\documentclass[fleqn,usenatbib]{mnras}

\usepackage{newtxtext,newtxmath}

\usepackage[T1]{fontenc}

\DeclareRobustCommand{\VAN}[3]{#2}
\let\VANthebibliography\thebibliography
\def\thebibliography{\DeclareRobustCommand{\VAN}[3]{##3}\VANthebibliography}

\usepackage[encapsulated]{CJK}
\usepackage{ucs}
\usepackage[utf8x]{inputenc}

\usepackage{graphicx}	
\usepackage{amsmath}	
\usepackage{rotating}
\usepackage{comment}
\usepackage{float} 
\usepackage{natbib} 
\usepackage{xcolor}
\usepackage{rotating}
\usepackage{bm}
\usepackage{soul}
\usepackage{physics}
\usepackage[T1]{fontenc}

\newcommand{\cntextsc}[1]{\begin{CJK*}{UTF8}{gkai}#1\end{CJK*}}
\newcommand{\tctext}[1]{\begin{CJK}{UTF8}{bkai}#1\ignorespacesafterend\end{CJK}}

\newcommand{\proptosim}{\mathrel{\vcenter{
 \offinterlineskip\halign{\hfil$##$\cr
 \propto\cr\noalign{\kern2pt}\sim\cr\noalign{\kern-2pt}}}}}

\renewcommand{\d}{\mathrm{d}}

\renewcommand{\P}{\mathrm{P}}     

\title[Formation of Dust Rings and Gaps]{Formation of Dust Rings and Gaps in Non-ideal MHD Disks Through Meridional Gas Flows}

\author[X. Hu et al.]{
Xiao Hu (\cntextsc{胡晓}),$^{1}$\thanks{E-mail: xiao.hu.astro@gmail.com}
Zhi-Yun Li,$^{1}$
Zhaohuan Zhu (\tctext{朱照寰})$^{2}$
and Chao-Chin Yang (\tctext{楊朝欽})$^{2}$
\\
$^{1}$Department of Astronomy, University of Virginia, Charlottesville, VA 22904, USA\\
$^{2}$Department of Physics and Astronomy, University of
  Nevada, Las Vegas, 4505 South Maryland Parkway, Las Vegas,
  NV 89154-4002, USA\\
}

\date{Accepted XXX. Received YYY; in original form ZZZ}

\pubyear{2022}

\begin{document}
\label{firstpage}
\pagerange{\pageref{firstpage}--\pageref{lastpage}}
\maketitle

\begin{abstract}
Rings and gaps are commonly observed in the dust continuum emission of young stellar disks. Previous studies have shown that substructures naturally develop in the weakly ionized gas of magnetized, non-ideal MHD disks. The gas rings are expected to trap large mm/cm-sized grains through pressure gradient-induced radial dust-gas drift. Using 2D (axisymmetric) MHD simulations that include ambipolar diffusion and dust grains of three representative sizes (1~mm, 3.3~mm, and 1~cm), we show that the grains indeed tend to drift radially relative to the gas towards the centers of the gas rings, at speeds much higher than in a smooth disk because of steeper pressure gradients. However, their spatial distribution is primarily controlled by meridional gas motions, which are typically much faster than the dust-gas drift. In particular, the grains that have settled near the midplane are carried rapidly inwards by a fast accretion stream to the inner edges of the gas rings, where they are lifted up by the gas flows diverted away from the midplane by a strong poloidal magnetic field. The flow pattern in our simulation provides an attractive explanation for the meridional flows recently inferred in HD 163296 and other disks, including both ``collapsing'' regions where the gas near the disk surface converges towards the midplane and a disk wind. Our study highlights the prevalence of the potentially observable meridional flows associated with the gas substructure formation in non-ideal MHD disks and their crucial role in generating rings and gaps in dust.
\end{abstract}




\section{Introduction}
\label{sec:intro}

Recent observations with the Atacama Large Millimeter/submillimeter Array (ALMA) have shown that protoplanetary disks generally do not have smooth dust distributions. They often display bright rings and dark gaps \citep[e.g.,][]{2015ApJ...808L...3A,2018ApJ...869L..42H,2020ARA&A..58..483A}. Planet-disk interaction is a widely discussed explanation of these substructures (see recent review by J. Bae et al. 2022, PPVII). The filtration effect on larger dust particles can make dust gaps more prominent than gas structure \citep[e.g.,][]{2012ApJ...755....6Z,2018ApJ...869L..47Z}. The gap features in HL Tau can be either explained by multi planets with one in each gap \citep[e.g.,][]{2015ApJ...809...93D,2015MNRAS.453L..73D,Dong2017}, or a single planet opening multiple gaps \citep{2018ApJ...864L..26B}.
And in a few cases, accurate measurements of
rotation curves revealed perturbed velocity patterns that
could be driven by Jupiter-mass planets embedded in the gas
disk \citep[e.g.,][]{2018ApJ...860L..12T,2019Natur.574..378T}. However, rare detection of
embedded planets in disks \citep[e.g.,][]{2013ApJ...766L...1Q,2015Natur.527..342S} leaves open other possibilities for the origin of the disk substructures.

Alternative mechanisms have been proposed to explain these features. For example, \citet{2015ApJ...806L...7Z} proposed that the condensation fronts of major volatiles (such as $\mathrm{H_2O}$ and $\mathrm{NH_3}$) can alter dust growth, which can lead to ring and gap formation, especially when the sintering effect between dust aggregates is taken into account \citep{2016ApJ...821...82O}. \citet{2019ApJ...885...36H} found that the snow line-induced changes in the dust distribution and ionization level lead to sharp changes in the magnetic diffusivity \citep{2016ApJ...821...82O}, which, in turn, lead to a spatially varying mass accretion rate that naturally produces gaseous rings and gaps.

Dust radial transport in protoplanetary disks has been extensively studied with one dimensional calculations \citep[e.g.,][]{2010A&A...513A..79B,2021A&A...647A..15D}. Most of such calculations adopted disk models with smooth surface density profiles. Long term particle radial drift in smooth disks inevitably meets the ``radial-drift barrier''. In order to explain dust particle retention in the outer disk, \citet{2012A&A...538A.114P} introduced sinusoidal perturbations in the disk's surface density, but did not include gas dynamics within the substructures. The primary goal of our investigation is to study how the dust is transported in global non-ideal MHD simulations where the gas substructures develop naturally. 

The validity of one dimensional dust transport has been tested using three-dimensional global unstratified magnetohydrodynamic (MHD) simulations including Lagrangian dust particles \citep{2015ApJ...801...81Z}. Unstratified disks work well for larger particles that settle efficiently to the midplane, but lack vertical structure for smaller particles that are prone to turbulence stirring. \citet{2016A&A...590A..17R} generated structures similar to inner region of HL Tau using a 3D MHD global disk simulation with Ohmic resistivity and multiple sized particles. Recently \citet{2020A&A...639A..95R} studied the settling and dynamics of smaller grains in an  ambipolar-dominated flow using global disk simulations, with the dust approximated as a fluid. These works illustrated several aspects of the dust behaviors in protoplanetary disks, such as radial concentration in the active zone, dust scale heights of different sizes and turbulence level, but leave open the role of gas flows, particularly the disk meridional motions, in shaping the dust distribution. How the gas motions affect the formation of dust substructure in a magnetized disk is the focus of our investigation.

This paper is organized as follows. In Section~\ref{sec:method}, we
describe the simulation setup, including the disk model, magnetic field, the boundary conditions and initialization of dust particles. The results of a fiducial simulation are presented in Section~\ref{sec:fid}, where we show that prominent rings and gaps are formed in both gas and dust, and the dust spatial distribution is mostly shaped by meridional gas flows associated with the gas substructure formation. In Section~\ref{sec:parameter}, we explore how changes in the magnetic field and the strength of ambipolar diffusion (AD hereafter) modify the picture of the fiducial run. In Section~\ref{sec:discussion}, we connect the simulation results with observations, proposing a new interpretation for the recently inferred meridional flows from ALMA observations. Finally, Section~\ref{sec:conclusion} concludes with the main
results of this study.

\section{Method}
\label{sec:method}
We solve the magnetohydrodynamic (MHD) equations in spherical-polar coordinates $(r, \theta, \phi=0)$ using Athena++ \citep{2020ApJS..249....4S} with non-ideal magnetic diffusion terms \citep{2017ApJ...836...46B,2017ApJ...845...75B}. Its MHD algorithms are based on (1) unsplit higher-order Godunov methods, (2) an extension of the constrained transport (CT) algorithm to enforce the divergence-free magnetic field constraint, and (3) a variety of spatial reconstruction algorithms and approximate Riemann solvers. The non-ideal induction equation with only ambipolar diffusion is:
\begin{eqnarray}
\frac{\partial {\bm B}}{\partial t}=\nabla \times \left({\bm v}\times {\bm B}\right)
-\frac{4\pi}{c}\nabla \times \left( 
\eta_\mathrm{A} {\bm J}_{\bot}\right),\nonumber \\ 
\label{eq:induction}
\end{eqnarray}
where $\bm v$ is the gas velocity and $\bm B$ the magnetic field.  
$\bm J$ is current density vector, with $\bm J_{\bot}$ as the 
current component perpendicular to the magnetic field. 
The quantity $\eta_\mathrm{A}$ is the ambipolar diffusivity. We do not include Ohmic dissipation or the Hall effect, which is reasonable for outer disks (e.g., \citet{Armitage2019}). The Hall effect may become important on the scale of tens of AUs, but its treatment is much more difficult and will be postponed to a future investigation.

\subsection{Initial disk configuration}
The gas disk's initial setup is similar to \citet{2018MNRAS.477.1239S}. The simulation domain can be roughly divided into two parts: a cold, dense disk and a hot, low-density corona. The disk has a constant aspect ratio, $h/r=0.05$ at all radii. We limit the cold, dense portion to two scale heights above (and below) the midplane, i.e., polar angle $\theta$ ranges from $\pi/2 - \theta_0$ to $\pi/2 + \theta_0$, in which $\theta_0=\arctan{(2h/r)}$. The gas density and temperature in the disk midplane both follow a power law with index $p$ and $q$, respectively:
\begin{eqnarray}
\rho(r,\pi/2) = \rho_0(r/r_0)^{p}  \\ \nonumber 
T(r,\pi/2) = T_0(r/r_0)^{q} 
\label{eq:profile}
\end{eqnarray}
where $p=-1.5$, $q=-1$, $r_0$ is the radius of the inner boundary of the computational domain, and $\rho_0$ and $T_0$ are the density and temperature at $r_0=1$ ($1~{\rm au}$ in the real world).

We use a quick $\beta$ cooling scheme with a cooling timescale that is only $10^{-10}$ of the local orbital period, so the temperature profile is effectively fixed over time. In order not to have a sudden heat-up on the gas when entering the disk atmosphere, in the initial setup, we employed a smooth vertical profile that has a transition zone between the cold disk and the hot corona, different from \citet{2018MNRAS.477.1239S}:
\begin{equation}
T(r,\theta)=
\begin{cases}
T(r,\pi/2) & \text{if }  |\theta-\pi/2| < \theta_0 \\
T(r,\pi/2)\ exp[(|\theta-\pi/2|\\
\ -\theta_0)/\theta_0 \times\ln(160)]
  & \text{if } \theta_0 \leq |\theta-\pi/2| \leq 2\theta_0  \\
160\ T(r,\pi/2); & \text{if } |\theta-\pi/2| > 2\theta_0\\
\end{cases}\label{eq:T}
\end{equation}
The vertical density profile is generated based on hydrostatic equilibrium, i.e.,$v_r=v_\theta=0$. The initial azimuthal velocity, $v_\phi$ is calculated using force balance at vertical and radial directions \citep{2013MNRAS.435.2610N}. In this way, we created a smooth density and velocity profile connecting the disk and corona. The temperature jump between the disk and corona leads to a weak thermal wind, which is greatly enhanced by magnetic fields in our fiducial simulation.

\subsection{Magnetic field and diffusivity}

Poloidal magnetic fields are initialized with vector potential generalized from \citet{2007A&A...469..811Z}:
\begin{equation}
A_\phi(r, \theta) = \frac{2B_{z0}R_0}{4+p+q}\left(\frac{r\sin\theta}{r_0}\right)^{\frac{p+q}{2}+1}\
[1+(m\tan\theta)^{-2}]^{-\frac{5}{8}}
\end{equation}
where $p$, $q$ and $r_0$ are from Eq.\ref{eq:profile}, and $m$ is a parameter that specifies the degree that poloidal fields bend, with $m\rightarrow\infty$  giving a pure vertical field.  We chose $m=0.5$ the same as \citet{2017ApJ...836...46B}. The value of $B_{z0}$ is set by the initial plasma $\beta$, the ratio of the thermal to magnetic pressure, which is $10^3$ everywhere at the midplane for the reference run to be discussed below.

The ambipolar diffusion coefficient $\eta_{\rm A}$ is related to the dimensionless Elsasser number, $Am$, and the local field strength $B$ through:
\begin{equation}
    \eta_{\rm A} = \frac{1}{Am\ \Omega_K}\frac{B^2}{4\pi\rho}
    \label{eq:etaA}
\end{equation}
where $\Omega_K$ is the Keplerian angular speed. Following \citet{2018MNRAS.477.1239S}, we prescribe a power-law density dependence for the Elsasser number $Am$: 
\begin{equation}
    Am=Am_0\ f(\theta)\ (\rho/\rho_0)^{\alpha_{\rm AD}}\ \Omega_K^{-1}
    \label{eq:am}
\end{equation}
with $Am_0=0.25$, $\rho_0=1$, and $\alpha_{AD}=0.5$. The $\theta$ dependence function is given by
\begin{equation}
f(\theta)=
\begin{cases}
\exp\left(\frac{\cos^2(\theta+\theta_0)}{2(h/r)^2}\right)& \text{if }  \theta<\pi/2-\theta_0 \\
1 & \text{if } \pi/2-\theta_0<\theta<\pi/2+\theta_0 \\
\exp\left(\frac{\cos^2(\theta-\theta_0)}{2(h/r)^2}\right) & \text{if } \theta>\pi/2+\theta_0\\
\end{cases}\label{eq:func_theta}
\end{equation}
which ensures that the low density region outside the disk is well coupled to the magnetic field.

\subsection{Grid and boundary}

The equations are solved for $r \in [1, 316] $ au and $\theta  \in [0.05, \pi-0.05]$. The simulation domain has $80\times 96$ grid cells in $r,\theta$ directions at the root level. We use three levels of static mesh refinement towards the midplane, and each refined level has half the cell size of the previous level, so the disk scale height is resolved by about 12.6 grids at the finest level. The coverage of the finest grids is from 10 to 100 au radially, and about 2.5 scale heights ($\approx 0.13$ radians) above and below the midplane, that contains our region of interest without being limited by time step at the inner boundary.

We use modified outflow boundary conditions at both inner and outer radial boundaries. At the inner boundary, instead of copying gas density and pressure from the innermost active zone to ghost zones, we apply the same power-law used in the gas initialization to extend gas density and pressure to ghost zones. For gas velocity, the azimuthal velocity follows a Keplerian curve, and the other two components are copied from the innermost active zone while restricting mass flux from outside entering the domain. Reflective conditions are used for the $\theta$ boundaries.

\subsection{Dust particles}

The particle module in Athena++ employs a leapfrog scheme for integrating Lagrangian particle trajectories, including size dependence and the aerodynamic gas-dust drag term, triangular-shaped cloud interpolation to account for the drag of the gas on the dust (Yang in prep). In this paper, we focus on how the gas substructures affect the dust and include only the gas drag on the dust. We postpone a treatment including the backreation of dust on gas to a future investigation. For the fiducial case, we have 3 particle sizes: 1~mm, 3.3~mm and 1~cm. The disk surface density at 1 au is $\Sigma_0=500~{\rm g~cm^{-2}}$, and assuming a mean molecular mass of 2.4, the temperature is  $T_0=644~{\rm K}$. This gives a density $\rho_0=2.67\times10^{-10}~{\rm g~cm^{-3}}$, which yields a mean free path of $7.5~{\rm cm}$ at the inner most radius. We included both Epstein and Stokes drag law in our particle module. All particles are expected to be in the Epstein regime unless the simulation produces strong gas concentration that is higher than the innermost midplane, which is unlikely. The particles are initialized at a thin slab near disk midplane (0.02 radian above and below midplane) with local Keplerian velocity. Radially, local particle number density follows the same gradient of gas volume density, so the ratio $\Sigma_{\rm d}/\Sigma_{\rm g}$ is constant. The particles have a constant density in the $\theta$ dimension.

\section{The Fiducial Model}
\label{sec:fid}
\subsection{Disk substructure in the gas}

\begin{figure*}
    \centering
    \includegraphics[width=1.0\textwidth]{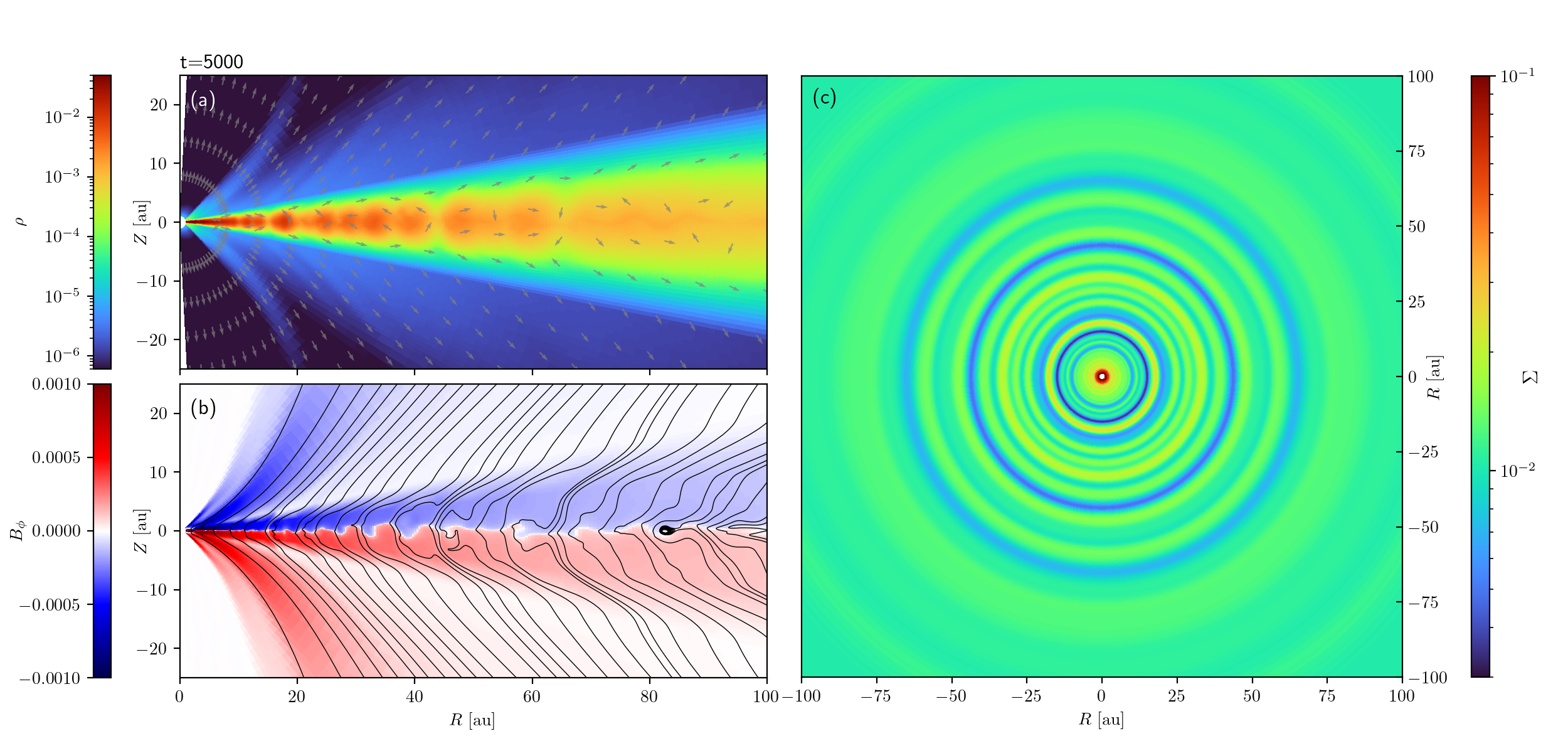}
    \caption{Overall gas structure of our fiducial model, plotted with the snapshot after 5000 innermost orbits (equivalent to 5000 years), in code units. Panel (a) and (b) are meridional views, and panel (c) is the face-on view. Panel (a) shows gas volume density with grey arrows indicating velocity vectors in the meridional plane, i.e., combination of $v_r$ and $v_\theta$. Panel (b) is the azimuthal magnetic field strength $B_\phi$, with poloidal field plotted as black solid lines. The field lines are plotted such that they have a constant magnetic flux between them at the midplane. The integrated column density $\Sigma$ is shown in panel (c), where rings and gaps can be seen clearly. (See the supplementary material in the online journal for an animated version of this figure.)
    }
    \label{fig:rhofluxbphi}
\end{figure*}

\begin{figure*}
    \centering
    \includegraphics[width=0.48\textwidth]{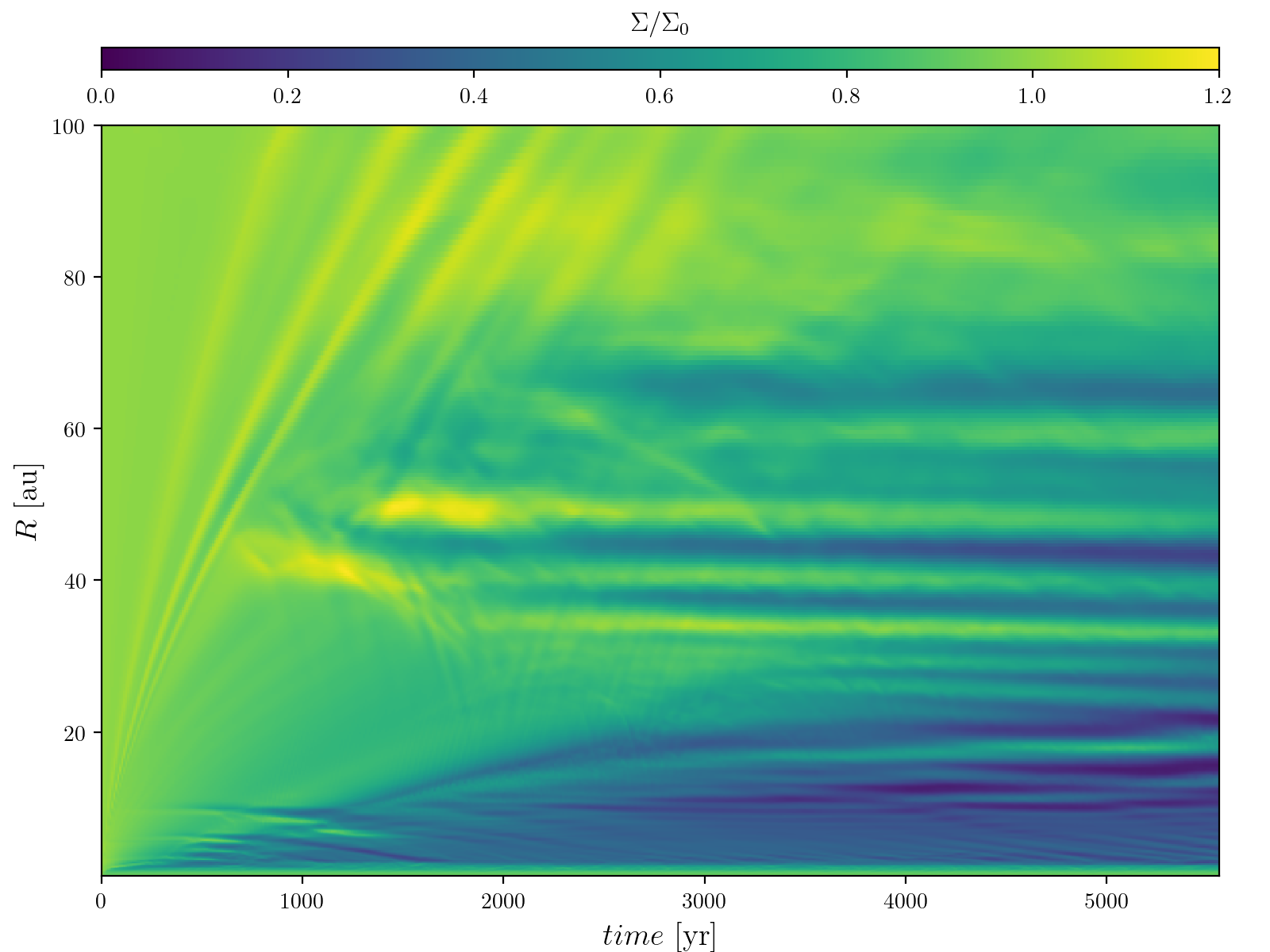}
    \includegraphics[width=0.48\textwidth]{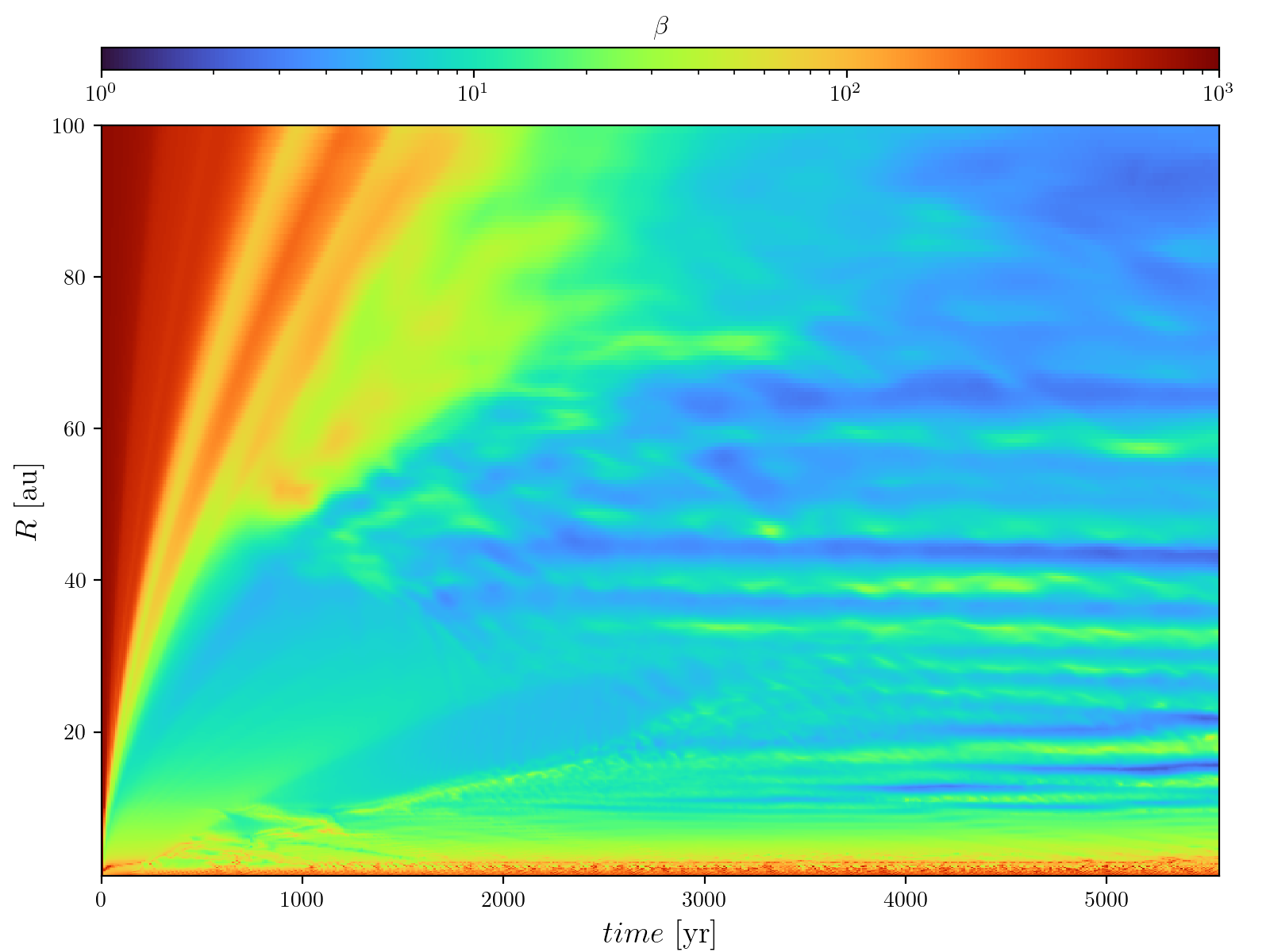}
    \caption{Time evolution of the gas disk's radial structure. {\it Left panel: }the gas surface density (normalized to its initial radial distribution); {\it right panel:} the plasma $\beta$, calculated using vertically averaged gas pressure and magnetic pressure within one scale height above and below the midplane. The development of stable rings and gaps persists for most of the simulation duration.}
    \label{fig:sigma_time}
\end{figure*}

In the fiducial model, prominent rings and gaps start to form after a few hundred times the period at the inner edge of the disk $t_0=1$~year. They become well established after a few thousand inner orbits. A snapshot of gas and magnetic field structure is presented in Fig.~\ref{fig:rhofluxbphi}. The gas structure is fairly stable at a representative time $t=5000\ t_0$ and changes relatively little at late times, as seen in the time evolution plots of the surface density and the plasma-$\beta$ (averaged over one disk scale height above and below the midplane) shown in Fig.~\ref{fig:sigma_time}. Their formation is broadly similar to that described in \citet{2018MNRAS.477.1239S}. Specifically, a toroidal magnetic field $B_\phi$ quickly develops in the wind-launching disk that is initially threaded by a purely poloidal magnetic field. The toroidal field reverses direction sharply near the midplane (see Fig.~\ref{fig:rhofluxbphi}b), producing a large magnetic tension force in the negative $\phi-$direction that acts to drain angular momentum from the field reversal region, producing a fast mid-plane accretion stream that is essential for both the gas and dust dynamics, as discussed in more detail below. The rapid accretion in the stream drags the poloidal magnetic field into a highly pinched configuration where the radial component $B_r$ changes direction sharply. The region of sharp field reversal (in both $B_\phi$ and $B_r$) appears highly dynamic. It can migrate to the disk surface before moving back to the midplane (see the movie version of Fig.~\ref{fig:rhofluxbphi} in the supplementary materials). From the distribution of the poloidal magnetic field lines on the meridian plane, we find closed magnetic field loops, which are likely produced by the reconnection of highly pinched poloidal field lines (see, e.g., the midplane region slightly beyond 80 au in Fig.~\ref{fig:rhofluxbphi}b). There may be other processes at work that create the decrease of the poloidal magnetic field in some regions of the disk and the field concentration in others \citep[e.g.,][]{2014ApJ...796...31B,2017A&A...600A..75B,2019A&A...625A.108R,2020A&A...639A..95R,2021MNRAS.507.1106C}. In any case, regions of poloidal field concentration are evacuated by faster accretion and/or fast gas removal by outflow, creating gaps. Material accumulates in regions of weak poloidal field, creating rings. 

The formation of rings and gaps in our simulations is perhaps not too surprising since our initial conditions and non-ideal MHD effects (AD) are similar to those of \citet{2018MNRAS.477.1239S}. One difference is that we fixed the spatial distribution of the temperature to the initial one \citep[as done in][]{2019ApJ...885...36H} whereas \citet{2018MNRAS.477.1239S} adopted an adiabatic index of $\Gamma=1.01$ so that fluid parcels retain their initial temperatures as they move around. 

Despite this difference, the results are broadly similar, which adds support to the growing evidence that the formation of prominent substructures in non-ideal MHD gas disks is a robust phenomenon. The main question that we seek to address is: how do dust substructures form in such structured non-ideal MHD disks?

\subsection{Disk substructure in the dust}

Prominent dust features are formed the fiducial model, as illustrated by Fig.~\ref{fig:dust_faceon} (see the movie version in the supplementary materials) where we plot on a face-on wedge of 0.1 radian in angular width the column density distribution of the gas and the projected (onto the midplane) locations of grains of three representative sizes (1~mm, 3.3~mm, and 1~cm). We will refer to the dust features in such a face-on view as substructure because they are closer to the features viewed by the observer than those in the edge-on (meridional) view to be shown in Fig.~\ref{fig:dust_meridian} below. Alternating bands of high and low dust concentration start to develop at small radii after only a few hundred inner orbits\footnote{The exact appearance of the dust substructure may depend on the number of particles used in the simulation.}. They appear at larger radii at later times. The dust substructures become fully developed up to 100~au after a few thousand inner orbits. These are highly dynamic structures that move both inwards and outwards. Occasionally, some of them appear to merge together, while others appear to split apart. There are broad similarities between the substructures for different dust sizes, with many of the features appearing around the same time and location and moving in and out in a similar fashion. However, there are a few notable differences. Firstly, there are prominent dust concentrations at some radii for grains of one size but not the others. For example, a prominent concentration exists for 1~mm grains at $\sim 22$~au at the time shown in Fig.~\ref{fig:dust_faceon} but not for 3~mm and 1~cm grains. Secondly, the contrast between regions of high and low dust concentration appears higher for larger grains. In particular, the 1~cm dust appears to be concentrated into narrower rings, leaving the regions between the rings more empty of grains. Perhaps most intriguingly, the dust concentrations do not necessarily line up with the maxima in the gas column density distribution, which is surprising if the dust is primarily concentrated by the gas pressure gradient as generally expected. The offset is an indication that the dust concentration in our simulation is more complicated than the simplest expectation, as we explore in detail next.

\begin{figure*}
    \centering
    \includegraphics[width=1.\textwidth]{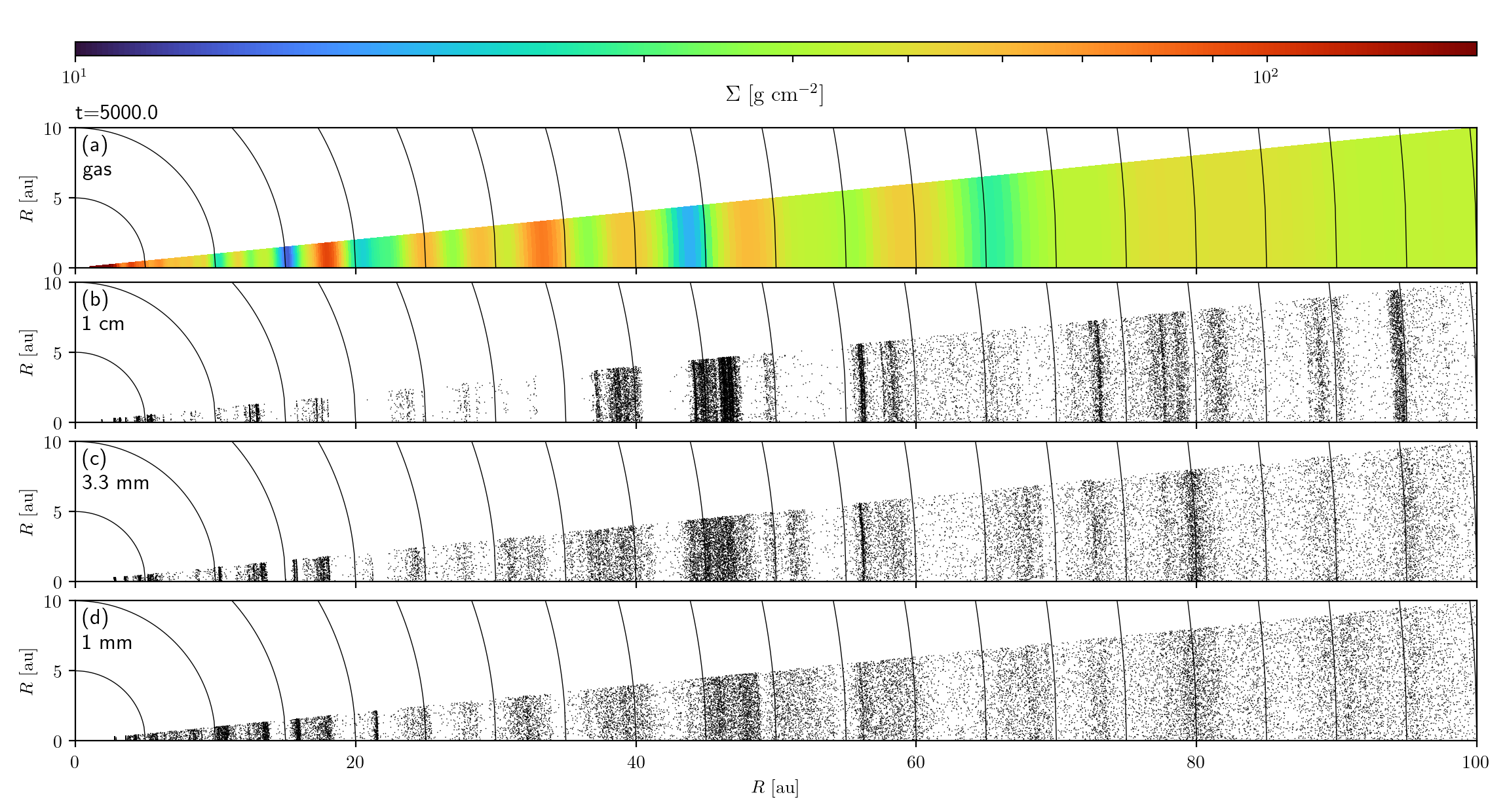}
    \caption{Face-on view of the gas and dust surface density distributions, showing prominent rings and gaps in both the gas (top panel) and grains of different sizes: panel (b), (c) and (d) are grains with size of 1~cm, 3.3~mm and 1~mm, respectively. To facilitate comparison, we plot a set of circles in steps of 5 au in radius (solid lines) in each panel. (See the supplementary material in the online journal for an animated version of this figure.)}
    \label{fig:dust_faceon}
\end{figure*}

A key to understanding the substructures in the face-on view of the dust surface density is the distribution of the dust on a meridional plane. The meridional dust distribution is shown in Fig.~\ref{fig:dust_meridian} and associated movie in the supplementary material.

\begin{figure*}
    \centering
    \includegraphics[width=1.\textwidth]{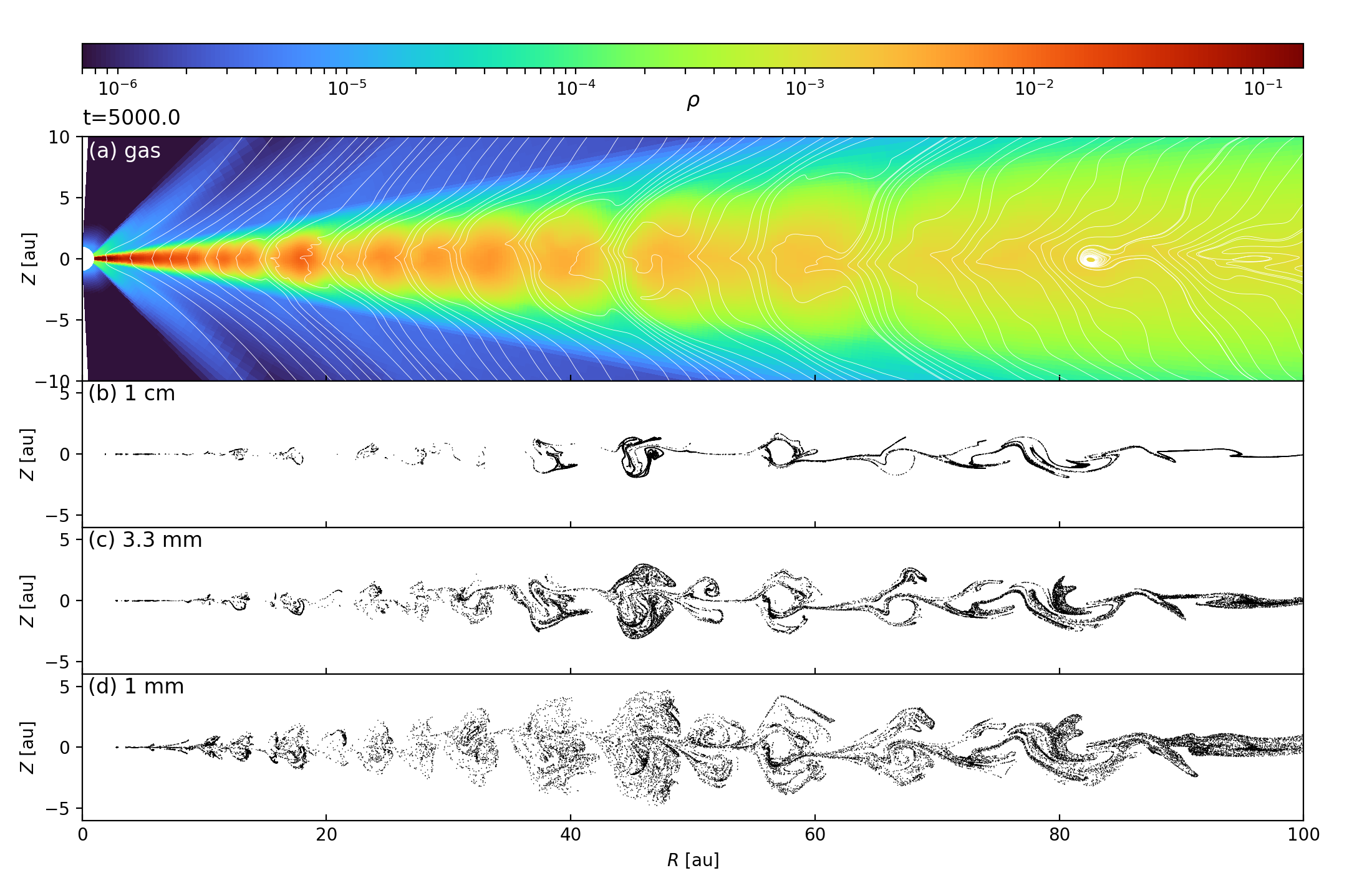}
    \caption{Meridional view of the gas and dust distributions, showing partially settled dust strongly stirred up by meridional gas flows. Panel (a) shows the gas density with the poloidal field illustrated as white solid lines. Panel (b), (c) and (d) are grains with sizes of 1~cm, 3.3~mm and 1~mm, respectively. (See the supplementary material in the online journal for an animated version of this figure.)}
    \label{fig:dust_meridian}
\end{figure*}

We begin the discussion of the dust distribution on a meridional plane with the simplest case of the largest, 1~cm, grains. As expected, such grains settle quickly towards the disk midplane, starting near the inner edge of the disk, and progress to increasingly large radii with time. The thin, settled, dust layer does not stay near the midplane, however, once prominent substructure develops in the gas. In particular, it is strongly stirred by meridional gas flows in the over-dense gas rings, where the gas circulation lifts the thin dust streams (or sheets in 3D) above and below the midplane against the (downward) vertical gravitational pull of the central star and often stretch and fold them \citep[similar stirring is also seen in other simulations, e.g.,][]{2018ApJ...868...27Y}. The thin dust streams that are strongly stirred up in both vertical and radial directions, when viewed from the face-on orientation, give rise to the rings and gaps in the dust surface density distribution. The dust rings in the face-on view tend to be much narrower than the gas rings, which are of course broadened by thermal pressure gradient that is negligible for the dust. Some of them arise from projection effects: for example, the ring at $r\sim 56$~au in the face-on view (see Fig.~\ref{fig:dust_faceon}b) is primarily caused by a longer path length through the inner edge of a loop-like dust structure in the meridional plane located between $\sim 55$~au and $\sim 60$~au (see Fig.~\ref{fig:dust_meridian}b). In other words, this ring is located where the (vertical) sight line passes through the (vertically extended) dust structure tangentially. As the dust streams are forced to circulate inside the rings by gas motions, they appear in the face-on view as substructures (particularly narrow rings) that oscillate back and forth in radius, especially at late times when the gas substructure has settled into a (statistically) quasi-steady state.  

The behaviors of the two smaller (3~mm and 1~mm) grains are broadly similar to that of the 1~cm grains. The main difference is that the smaller grains are less vertically settled before prominent gas substructures develop. As a result, the dust layer that is moved around by the meridional gas motions is thicker to begin with, which leads to, broadly speaking, a thicker version of the (highly inhomogeneous and dynamic) meridional structure that develops in the 1~cm dust. Because of this width difference and the tendency for larger grains to settle closer to the midplane, grains of different sizes are distributed differently in space and therefore experience somewhat different gas velocity fields at the dust locations. As a result, there are features in the distributions of small grains that are not found in larger grains. For example, there is a prominent elongated dust clump between 50 and 55~au that is vertically extended at $t=5000$~inner orbits for the 1~mm and 3.3~mm grains, which has no obvious counterparts for the larger 1~cm grains (see Fig.~\ref{fig:dust_meridian}). This dust clump produces a ring in the surface density distribution of the 3.3~mm dust that does not exist in the larger grains (see Fig.~\ref{fig:dust_faceon}). Similarly, in Fig.~\ref{fig:dust_faceon}, we see a relatively wide band of high dust concentration near 45~au for the 1~cm grains that does not show up as clearly in the surface density maps of the smaller 1~mm and 3.3~mm grains. Again, such differences are expected because grains of different sizes are spatially distributed differently and thus sample different disk physical conditions, especially density and velocity fields. 

\subsection{Dust radial transport: dust-gas drift and  sub-/super-Keplerian regions}

The dust moves in the disk at a velocity that can be decomposed into 
\begin{equation}
    \mathbf{v}_\mathrm{dust}=\mathbf{v}_\mathrm{gas}+(\mathbf{v}_\mathrm{dust}-\mathbf{v}_\mathrm{gas})\equiv\mathbf{v}_\mathrm{gas}+\Delta\mathbf{v}
    \label{eq:v_dust}
\end{equation}
where $\Delta\mathbf{v}=\mathbf{v}_\mathrm{dust}-\mathbf{v}_\mathrm{gas}$ is the dust drift velocity relative to the gas. This expression points two distinct mechanisms to move dust around in a disk and concentrate it to localized regions: through either $\mathbf{v}_\mathrm{gas}$ (advection by gas) or $\Delta\mathbf{v}$ (dust-gas drift). The latter includes the  conventional mechanism for dust trapping through a gas pressure bump, which creates super-Keplerian rotation interior to the pressure maximum and sub-Keplerian rotation exterior to it, which, in turn, causes the dust to drift radially relative to the gas towards the pressure maximum through angular momentum exchange between the dust and gas via aerodynamic drag \citep{fW72}. The less widely appreciated mechanism is the trapping of the dust by gas advection (through $\mathbf{v}_\mathrm{gas}$; see J. Bae et al. 2022, PPVII, for a recent review). In this subsection, we will show that conditions for the radial drift-driven dust concentration exist in our simulated disk, but this mechanism is overshadowed by that from the more dynamic gas advection. We should note that the disk in the reference model is more dynamically active and accretes at a higher rate than a typical Class II protoplanetary disk. It is more representative of younger (e.g., Class 0) disks (see \S\ref{sec:discussion} below). 

\begin{figure*}
    \centering
    \includegraphics[width=1.0\textwidth]{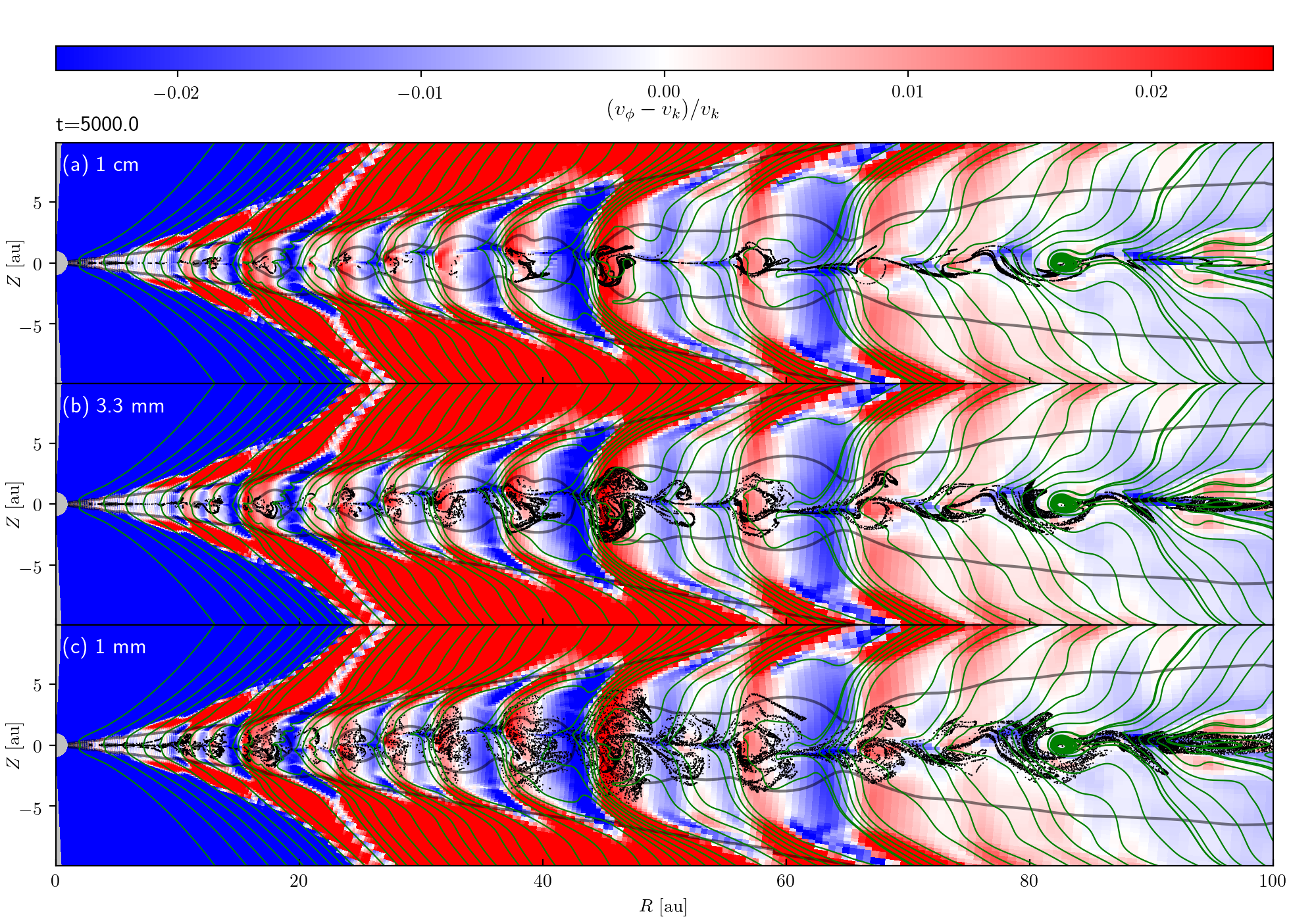}
    \caption{Fractional difference between the gas rotation speed and the local Keplerian speed, showing alternating super-Keplerian (red) and sub-Keplerian (blue) regions at $t=5000$~years. The boundaries between the super-Keplerian regions and their adjacent exterior sub-Keplerian regions are expected to serve as traps for large grains through dust-gas radial drift. Note the magnetic braking-induced sub-Keplerian rotation in corrugated current sheet near the disk midplane where the toroidal magnetic field changes sign. The black lines are contours of constant $\rho r^{-1.5}$ that indicate the locations of rings and gaps. Poloidal magnetic fields are plotted as green lines. Plotted in black dots are the locations of grains of 1 cm (panel a), 3.3 mm (b) and 1 mm (c). Note that while most of the vertically extended largest (1~cm) grains are  concentrated in the (red) super-Keplerian regions, some of their smaller counterparts extend well into the (blue) sub-Keplerian regions. (See the supplementary material in the online journal for an animated version of this figure.)}
    \label{fig:sub-super-K}
\end{figure*}

To show the potential for drift-driven dust concentration, we plot in Fig.~\ref{fig:sub-super-K} the fractional difference of the gas rotation speed and the local Keplerian speed at the representative time $t=5000$~years. Clearly, there are coherent, alternating regions of super-Keplerian (red) and sub-Keplerian (blue) rotation on the disk\footnote{Such alternating super-Keplerian and sub-Keplerian regions should also exist in other comparable non-ideal MHD disk studies, such as \cite{2018MNRAS.477.1239S,2020A&A...639A..95R,2021MNRAS.507.1106C}, where gas rings and gaps are formed.}. Some of these regions extend into the wind. This is particularly true inside $\sim 50$~au, where the locally super-Keplerian zones start to merge together into a nearly contiguous super-Keplerian wind away from the disk. The super-Keplerian rotation in the wind zone is to be expected for a magnetically driven wind, since it is receiving angular momentum from the magnetically braked material inside the disk. As an example, we note that, in the classic magnetocentrifugal wind of \citet{1982MNRAS.199..883B}, fluid parcels are flung out along rigid inclined magnetic field lines as ``beads on a wire'' because of super-Keplerian rotation. 

Inside the disk where the thermal pressure dominates the magnetic pressure (i.e., with a plasma-$\beta$ much larger than unity, particularly inside the dense gas rings; see Fig.~\ref{fig:sigma_time}b), the deviation from Keplerian rotation is caused mostly by the thermal pressure gradient.  The ``attractive'' Keplerian lines that divide the super-Keplerian disk regions from their adjacent {\it exterior} sub-Keplerian neighbors are located close to the pressure maxima inside high density gas rings. Grains are expected to drift (relative to the gas) towards these locations. Conversely, the ``repulsive'' Keplerian lines that divide the super-Keplerian regions from their adjacent {\it interior} sub-Keplerian regions tend to locate near the inner edges of gas rings. Grains are expected to drift away from these locations.

Magnetically induced deviation of gas rotation from the local Keplerian value is not limited to the wind zone. This can be seen most clearly at early times in the movie version of Fig.~\ref{fig:sub-super-K} in the supplementary material, when classic ``S-shaped'' channel flow-like features \citep{1992ApJ...400..595H} are prominent, particularly in the wind-disk transition zone near the disk surface. 

Angular momentum is removed magnetically from the inner parts of such (unstable) features, which rotate at a sub-Keplerian speed. It is magnetically transferred to the outer parts of the features, which are forced to rotate at a super-Keplerian speed. At later times when rings and gaps are well developed, the magnetic braking-induced sub-Keplerian rotation can be seen most clearly in the corrugated current sheet where the toroidal magnetic field $B_\phi$ changes sign (Panel (b) in Fig.~\ref{fig:rhofluxbphi}). It is particularly evident in the low-density gaps where the plasma-$\beta$ is relatively low (see Fig.~\ref{fig:sigma_time}b) and the poloidal magnetic field lines are concentrated, which lead to efficient angular momentum transfer along field lines from near the midplane to the disk surface and beyond, where the gas tends to have a super-Keplerian rotation.

\begin{figure*}
    \centering
    \includegraphics[width=1.\textwidth]{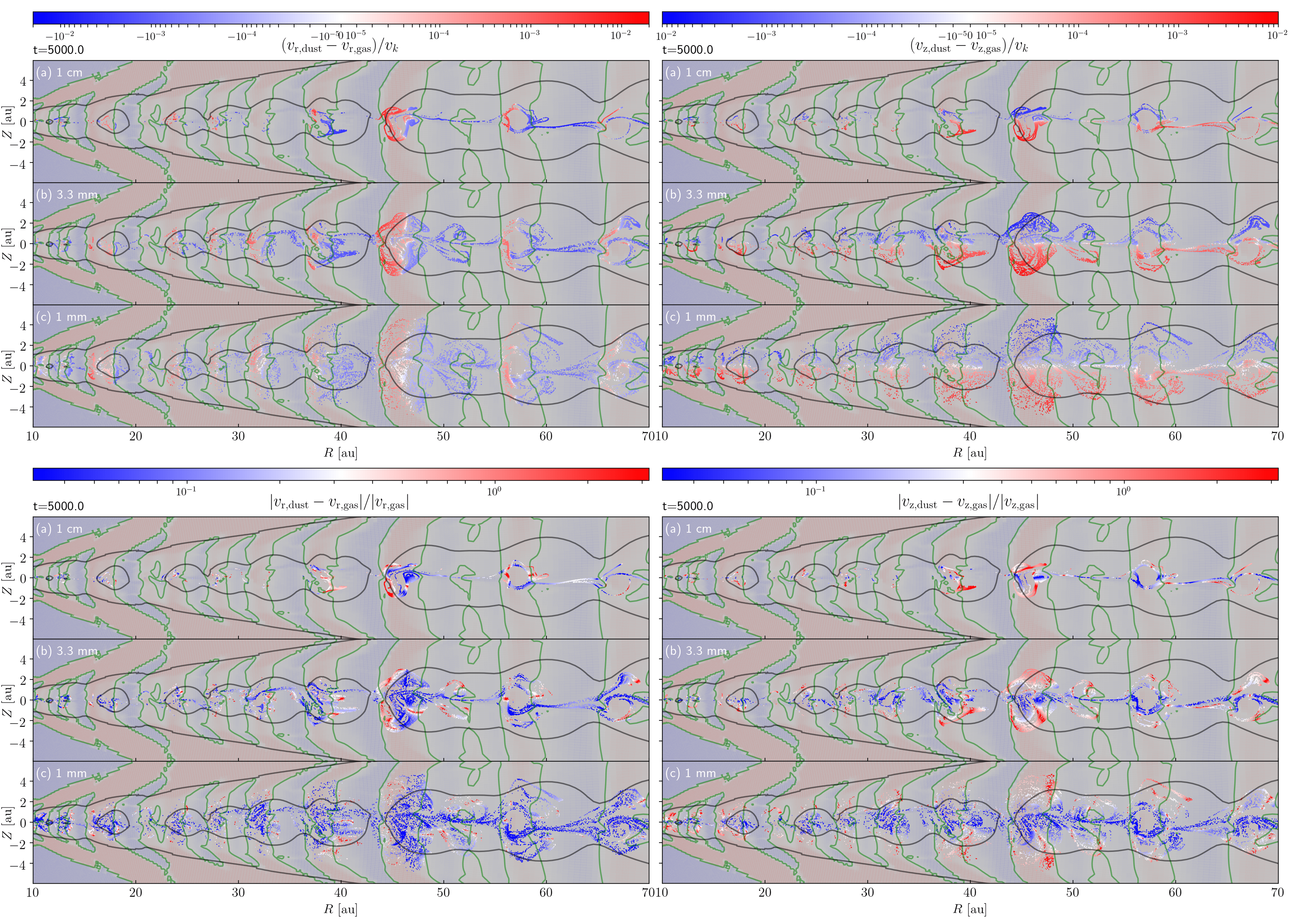}
    \caption{{\bf Upper left:} Radial drift velocity of the dust relative to the gas normalized by the local Keplerian speed, over plotted on the map of the ratio of the gas rotation speed and the local Keplerian speed that shows sub- and super-Keplerian rotation. {\bf Lower left:} Ratio of the dust radial drift speed and the local gas radial speed. The green contours mark the location where gas is rotating just at Keplerian velocity, The black lines are contours of constant $\rho r^{-1.5}$ that indicate the locations of rings and gaps. {\bf Upper right:} Vertical drift velocity of the dust relative to the gas normalized by the local Keplerian speed, over plotted on the map of the ratio of the gas rotation speed and the local Keplerian speed that shows sub- and super-Keplerian rotation. {\bf Lower right:} Ratio of the dust vertical drift speed and the local gas vertical speed. 
    }
    \label{fig:vrvz_drift}
\end{figure*}

The presence of alternating sub-Keplerian and super-Keplerian regions should be conducive to dust concentration through radial drift. This expectation is broadly met, as illustrated in the upper left panels of Fig.~\ref{fig:vrvz_drift}, which shows the radial drift velocity of the dust relative to the gas normalized by the local Keplerian speed for grains of all three sizes at the representative time $t=5,000$~years. The majority of the dust particles in the super-Keplerian regions are indeed drifting outward (relative to the gas) and those in the sub-Keplerian regions are drifting inward, as expected because they tend to experience a tail-wind in the former and a head-wind in the latter (see, e.g., the ring at $\sim 50$~au). However, this is not strictly true for all particles. In particular, some of the particles in the super-Keplerian region of the ring located $\sim 40$~au are drifting inward, which is an indication that the traditional picture does not fully capture the complexities of the radial drift in our highly dynamic and magnetized system. 

\subsection{Dust radial transport: advection by gas motions}

Although dust radial drift induced by non-Keplerian gas rotation exists, it is unlikely to be the dominant mechanism for the dust concentration observed in our simulation. This is supported by the fact that most of the grains are concentrated in the inner parts of the gas rings, to the left of the ``attractive'' Keplerian lines where the concentration is expected in the traditional picture. The offset is particularly true for the largest grains, which are more spatially confined. It is evidence that another mechanism for dust transport must be at play. We show below that it is the advection by the gas motion.

One of the regions where the gas advection of dust can be seen most clearly is the narrow midplane current sheet (see Fig.~\ref{fig:rhofluxbphi}b for an illustration), where the dust is advected inwards by the rapid gas accretion caused by strong magnetic braking. This is true even in the inner parts of dense rings where the gas is rotating at a super-Keplerian speed, which tends to force the dust to move radially outward relative to the gas. Once the grains are advected close to the inner edge of a dense ring, most of them do not exit the ring (and into the gap); rather, they are lifted up against the vertical gravitational force of the central star by a meridional gas circulation that confines the dust mostly to the inner half of the ring. The circulation pattern is likely driven by the narrow fast accreting stream near the midplane current sheet, part of which is diverted away from the stream because its inward motion is blocked, at least in part, by the concentrated poloidal magnetic field lines that drape around the inner boundary of the ring (see the top panel of Fig.~\ref{fig:dust_meridian} for an illustration). The strong, largely stationary, poloidal field forms a barrier to the gas (and the dust trapped by it), which can cross the field lines only through non-ideal MHD effects (in this case, ambipolar diffusion) and magnetic reconnection. The exact role of the accretion stream in dust transport will be discussed further below.

To evaluate the relative importance of dust-gas drift and gas advection in the radial dust concentration more quantitatively, we plot in the lower left panels of Fig.~\ref{fig:vrvz_drift} the absolute values of the ratio of the radial drift speed and gas speed. It is clear that most grains drift at a small fraction of the gas speed, even for the largest dust. Notable exceptions include the large particles near the inner edge of the ring near 50 au, where the midplane gas (inward) accretion stream is diverted away from the midplane and starts to circulate outward, with a relatively small local radial speed that is comparable to, or even less than the dust radial drift speed. In such (limited) regions, the drift dominates the gas advection in moving the dust around radially, because the radial motion of the gas ceases temporally as it turns around from inward accretion to outward expansion. 

\subsection{Dust vertical transport: meridional circulation and gravitational settling}
\label{sec:vertical}

The competition between the gas advection and dust-gas drift determines the vertical dust transport as well. Because of the unbalanced vertical gravitational pull of the central star towards the disk midplane, the dust particles will always have a tendency to drift relative to the gas toward the midplane. This expected sedimentation is shown in the upper right panels of Fig.~\ref{fig:vrvz_drift}, where the vertical component of the drift velocity is plotted for each of the particles in the simulation. For particles of a given size, the downward drift velocity tends to be larger at a larger distance from the midplane \citep[see e.g.,][Fig.~2]{YJ16}, because the vertical gravity is stronger and the density is lower, both of which are conducive to the vertical dust settling. Nevertheless, the dust is far from being completely settled to midplane, even for the largest dust grains. This is because the tendency for the gravitational settling is constantly being countered by gas updrafts, which lift the dust particles up before they can settle to and stay near the midplane. 

To show the competition between the gas advection and dust-gas drift more quantitatively, we plot in the lower right panels of Fig.~\ref{fig:vrvz_drift} the ratio of the drift velocity and the gas velocity in the vertical direction. Clearly, the gas has a much higher vertical velocity than the dust settling velocity in regions not far from the midplane for grains of all three sizes (i.e., the particles with blue and white colors), indicating that the low-altitude gas current is capable of  moving the dust around, particularly away from the midplane. However, at higher altitudes, the dust settling velocity starts to dominate the gas vertical velocity (i.e., the particles with red colors), indicating that the dust begins to decouple from the gas and starts to settle towards the midplane. Not surprisingly, the larger grains start to decouple from the gas at a lower altitude, whereas the smaller grains can be carried by the gas currents further away from the midplane, which gives them a puffier appearance. 

We conclude that it is the highly dynamic gas currents in the disk that dominate the dust dynamics, transport, and distribution in both the radial and vertical directions, although dust radial drift and vertical settling play a role as well. In the next subsection, we will examine the competition between gas advection and dust-gas drift from another perspective, focusing on the strong (warped) midplane
gas current.

\subsection{Fast mid-plane accretion stream and meridional circulation}

Since the gas currents in the disk lie at the heart of the dust transport and concentration, we have decided to illustrate these currents more visually with line integral convolution (LIC) to show the gas velocity field in the meridional plane.

\begin{figure*}
    \centering
    \includegraphics[width=1.\textwidth]{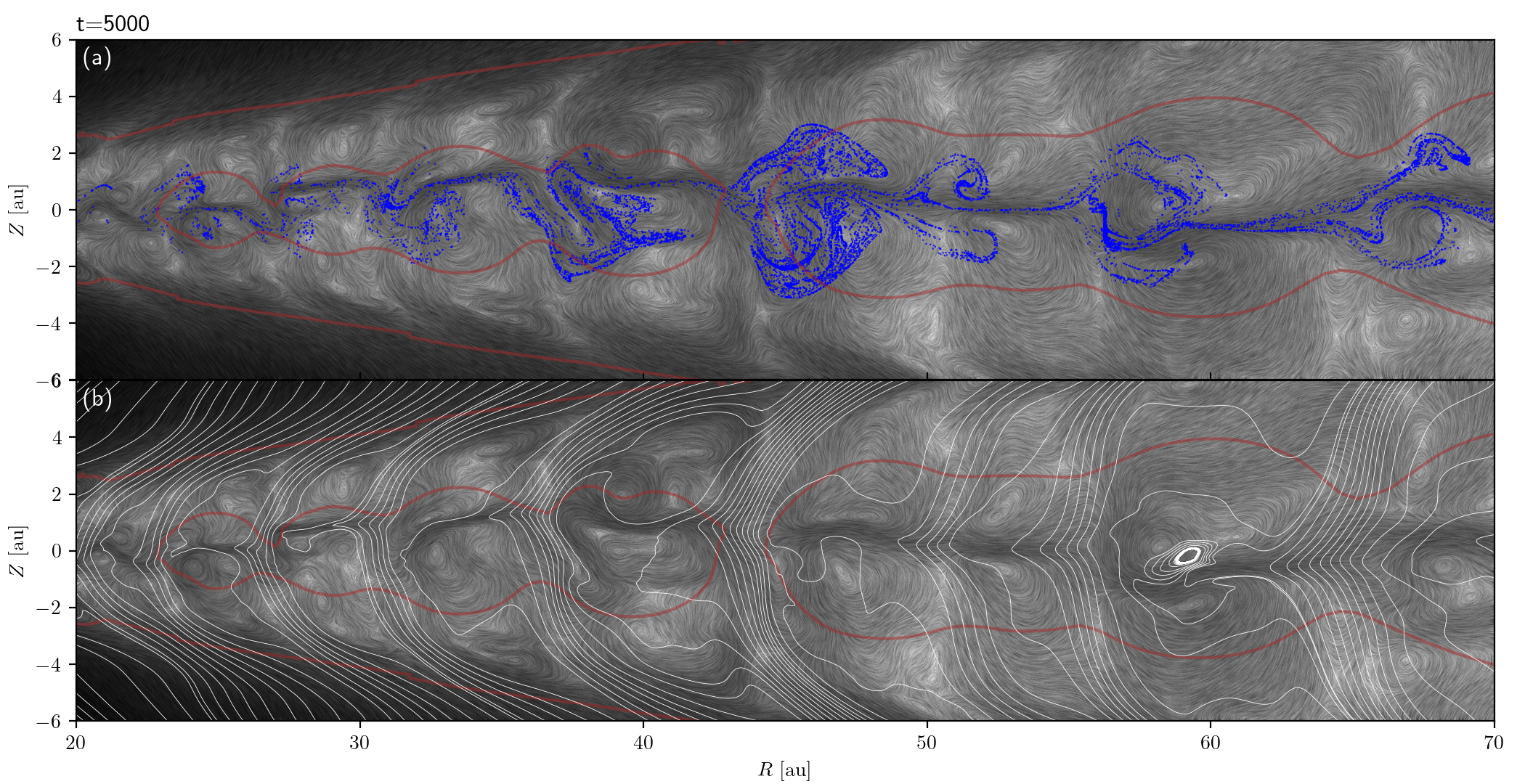}

   \caption{{\bf Top panel:} 3.3~mm dust particles superposed on LIC (line integral convolution) map of the gas velocity field in the meridian plane for a representative time $t=5000$~years. The darker the color, the faster the gas moves. The dark red lines are iso-contours of the density normalized by the initial midplane density, highlighting the locations of the gas rings. 
   {\bf Bottom panel:} magnetic field lines superposed on the LIC map. (See the supplementary material in the online journal for an animated version of the top panel with three particle sizes. )}
    \label{fig:LIC}
\end{figure*}

Fig.~\ref{fig:LIC} and the associated movie show the velocity field of the gas in the disk clearly. Broadly speaking, there are three distinct components to the velocity field: midplane accretion stream, field-crossing accretion funnel, and back circulation. The midplane accretion stream is the dominant gas kinematic feature throughout the disk, particularly inside the dense rings. As discussed earlier, it is produced by the strong magnetic braking associated with the sharp kink in the toroidal magnetic field where its polarity reverses (see Fig.~\ref{fig:rhofluxbphi}c for an example). The fast midplane accretion stream in a ring encounters the more or less stationary, strong bundle of poloidal magnetic field lines that drape around the inner boundary of the ring (see Fig.~\ref{fig:sub-super-K} and \ref{fig:LIC}). It cannot continue unimpeded because doing so would lead to a strong pinch of the (strong) poloidal field and a large outward magnetic tension force that would slow down the inward motion. The damming by the concentrated poloidal field lines forces the fast midplane accretion stream to divert away from the midplane, where it bifurcates into two separate streams. One stream is the field-cross accretion funnel, where a fraction of the diverted gas crosses the poloidal field lines (through ion-neutral drift in the presence of ambipolar diffusion) at a relatively large distance from the midplane. Once loaded onto the poloidal field, the gas tends to slide along the (pinched) field lines towards the midplane, forming a funnel-shaped collection of streamlines (see the streamlines between $\sim 42$~au and $\sim 45$~au in Fig.~\ref{fig:rhostreamzoom} below). One part of the remaining fraction of the gas diverted away from the midplane is forced to move backward (i.e., outward) before converging back toward the midplane and rejoining the midplane accretion streaming, forming a back circulation (see the green streamlines between $\sim 47$~au and $\sim 50$~au in Fig.~\ref{fig:rhostreamzoom} below); the other part leaves the disk to join the wind. 
We should note that all these kinematic components are highly dynamic and may not always be cleanly identifiable. In particular, the back (meridional) circulation may contain more than one vertices, with complex morphology and dynamics. Nevertheless, they provide a broad framework to interpret the dust dynamics and distribution seen in our simulation. These distinct kinematic features are illustrated further in Fig.~\ref{fig:rhostreamzoom} below. 

We will first focus on the dust of intermediate size ($3.3$~mm, Panel (a) of Fig.~\ref{fig:LIC}), which has two very distinct types of vertical dust distributions. In the low-density strong-poloidal field gaps and the outer halves of the dense rings, the dust tends to be concentrated in a narrow stream near the midplane. This is not surprising given the gas flow patterns in these two regions. Specifically, in a gap region, the field-crossing accretion funnel tends to channel the dust towards the midplane in the gap region, where the dust becomes trapped by the fast gas accretion stream (see the streamlines around radius $r=42$~au in Fig.~\ref{fig:rhostreamzoom} below). In the outer half of a ring, the back-circulation gas current keeps the dust entering the ring from the neighboring gap in the narrow midplane accretion stream. This is in contrast with the dust distributions in the inner halves of the rings, which are much more vertically extended, because the gas stream diverted away the midplane lifts the dust particles to larger vertical heights. Some of the lifted particles enter the gap with the field-crossing accretion funnel. The remaining ones are pushed outward by the back-circulation gas current. However, the outward-moving dust particles are not completely coupled to the circulating gas, especially at relatively large vertical heights, because the dust gravitational settling acceleration increases with height due to an increasing vertical gravitational force and a decreasing gas density (see right panels of Fig.~\ref{fig:vrvz_drift}). As a result, the dust is not coupled to the outermost part of the back circulation (which reaches the largest height and lowest density) that would have carried the dust to the back half of the ring. Instead, the particles cross gas streamlines, moving downwards as they are pushed outward, eventually becoming re-coupled to the denser, lower-height, inner back-circulation gas current which forces them to the midplane accretion stream before they can reach the outer edge of the gas ring (see also Figs.~\ref{fig:dust_faceon} and \ref{fig:dust_meridian}). This, we believe, is the reason why the vertically extended dust distribution is confined in the inner half of a ring.
\begin{figure*}
    \centering
    
\includegraphics[width=1.0\textwidth]{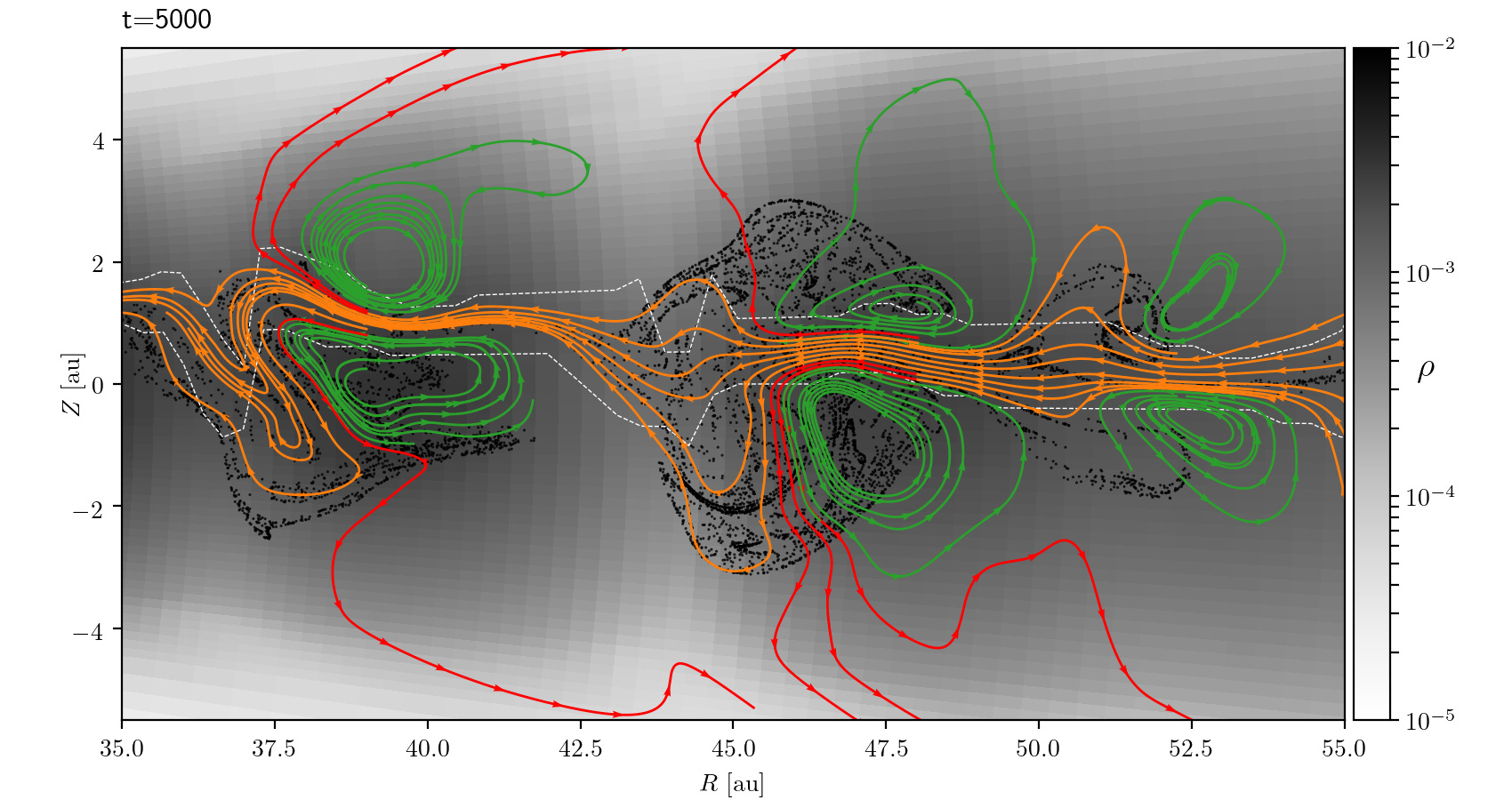}
    \caption{A zoomed-in view of 3.3~mm dust particle distribution with gas kinematic features. The white dashed lines mark the boundary of the fast midplane accretion stream. The solid lines are gas stream lines. The green lines clearly illustrate the meridional gas circulation patterns (backward or outward outside the midplane accretion stream) that are particularly prominent in dense gas rings. The ``flaring" orange stream lines around R=45~AU are an example of the field-crossing accretion funnel that moves radially cross the bundle of poloidal field lines in the gap (not shown) and into the next ring. Disk materials on the red stream lines are fed to the disk wind. A background black-and-white map for the gas density is plotted to highlight gas rings and gaps. 
    }
    \label{fig:rhostreamzoom}
\end{figure*}

The dust distributions of the larger (1~cm) and smaller (1~mm) grains can be broadly interpreted in a similar framework. The main difference is that the larger grains gravitationally settle more quickly toward the midplane, and thus has a narrower vertical extent. The faster settling also means that the vertically narrow part of the dust distribution follows the midplane accretion stream more closely and the extended part is more confined in the radial direction. The latter is because the larger dust is coupled to only the part of the back circulation gas current closest to the inner edge of the ring. Conversely, the smaller grains can be lifted to larger heights and carried further away from the inner edge of the ring by the back circulation gas current, which results in a larger extent in both vertical and radial directions. Nevertheless, even for the smallest grains under consideration in this simulation (1~mm), there is enough dust-gas drift that there is little dust reaching the outer edge of the ring, leaving the dust distribution concentrated toward the inner half. Local concentration should become weaker still for even smaller particles.

\subsection{Comparison with smooth viscous disks} 

\begin{figure}
    \centering
    \includegraphics[width=0.5\textwidth]{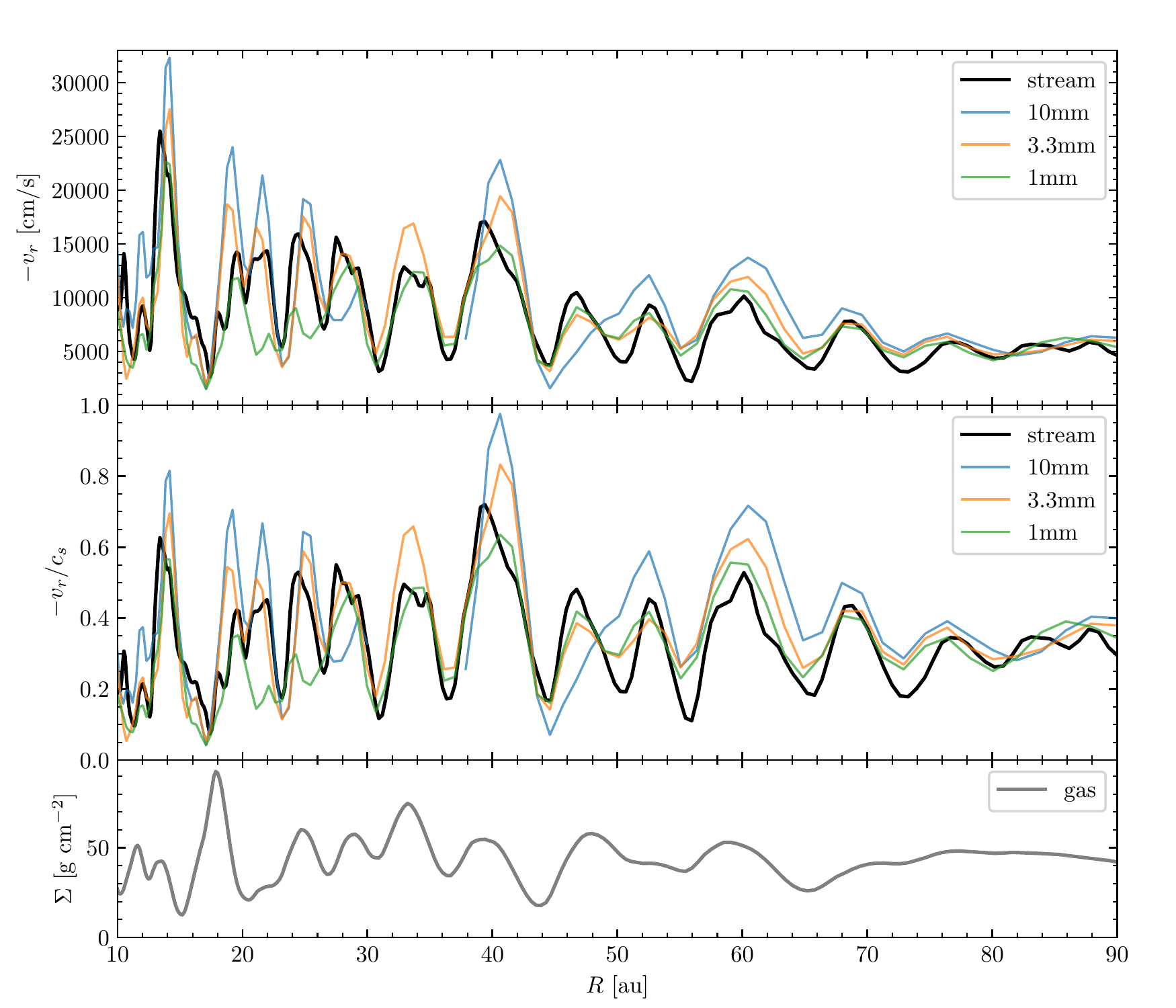}
    \caption{Radial velocity in the midplane accretion stream, with the gas plotted in solid black lines, and three particle sizes in color. The top panel shows absolute velocities, and the middle is the radial velocity scaled with local sound speed at midplane. In the bottom panel we plot the gas surface density to show the locations of rings and gaps relative to the radial velocity variations. }
    \label{fig:streamvrgasdust1d}
\end{figure}

\begin{figure}
    \centering
    \includegraphics[width=0.5\textwidth]{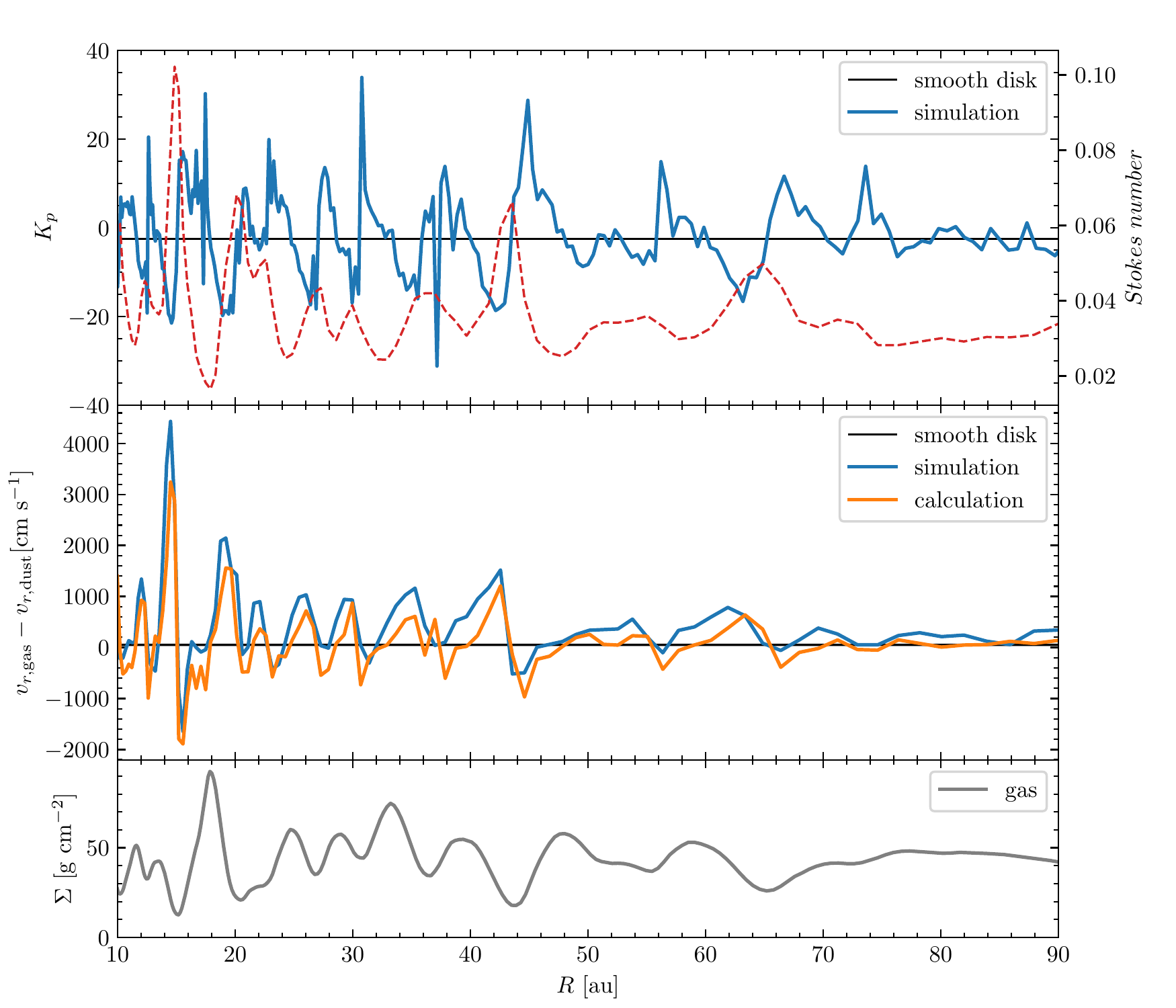}
    \caption{{\it Top:} Gas radial pressure gradient ($\d\ln{P}/d\ln{R}$) in the accretion stream (``simulation''), plotted together with the constant value of a smooth disk (black line). The red dashed line scaled by the right y-axis is the Stokes number of 3.3 mm particles. {\it Middle:} 3.3 mm particle drift velocity with respect to gas, with the blue line showing the measurement in the simulation and the orange line showing the calculation based on $K_P$ profile in the top panel. The black solid line is the drift velocity in a smooth disk. {\it Bottom:} gas surface density as a reference for location of rings and gaps.}
    \label{fig:kpvr_drift1d}
\end{figure}

In this subsection, we quantify the radial transport in the fast mid-plane accretion stream and compare it with smooth viscous disks. First, we define the spatial boundary of the midplane stream in the following manner: at each radius, the computational cell with maximum radial mass flux in gas is treated as the center of the stream. The upper and lower boundary of the stream is identified as where the radial mass flux drops by a factor of $e$. To ensure the profile of the stream is not affected by abrupt variations, we constrain the width of the stream to vary by no more than $20\%$ from one radius to the next. This constraint underestimates stream width where the accretion flux is ``puffed up'' by the strong poloidal field near each ring's inner boundary. The results are the white dashed lines in Fig.~\ref{fig:rhostreamzoom}. The particles are settling towards the stream, instead of the geometric midplane.
Later we'll show that the stream serves as the main driving force for particle's radial transport. 
Here we reconfirm the mechanism behind the width of dust rings: the better the coupling is, the longer the particles stay in the vertical and back (outward) circulation, and travel further out inside a gaseous ring, leading to a wider dust concentration.

To illustrate the gas flow structure in gaseous rings, we select vertical slices in the middle of the rings, and select several dots as starting location for streamline integration. The flow structure looks quite laminar in most part of the accretion stream, and much less so at the inner edge of gas rings, where most of the streamlines start to peel away from the stream. Those that are close to the center of the stream are able to ``penetrate'' through the gap and enter the next ring with relatively little change of direction. Other streamlines initially peel away from the stream before rejoining it in the gap (coded orange). These gap-crossing streamlines are part of the field-crossing accretion funnel discussed earlier. Some streamlines leave the disk entirely and enter into the disk-wind (coded red). The remaining streamlines loop back towards the midplane (coded green; they are part of the meridian back circulation discussed earlier).

The radial transport inside the stream is particularly effective compared to the traditional $\alpha$ disk model commonly used in dust evolution calculations. The gas radial velocity in a viscous accretion disk is \citep{LP74}:
\begin{equation}
    v_{r,g} = -\frac{3}{2}\frac{\nu}{r}
\end{equation}
where $\nu=\alpha c_s h$, with $c_s$ and $h$ being the sound speed and disk scale height respectively. With a constant aspect ratio of $h/r=0.05$, we can obtain a radial gas velocity:
\begin{equation}
    v_{r,g} = -2.35\times \alpha_{-3}r^{-0.5}_{\rm 10 au}\ \ {\rm cm~s^{-1}}
\end{equation}
where $\alpha_{-3}=\alpha/10^{-3}$, $r_{\rm 10 au}=r/{\rm 10 au}$. Comparing to the accretion speed in the stream shown in Fig.~\ref{fig:streamvrgasdust1d}, which is on the order of $10^4~{\rm cm~s^{-1}}$, this is a negligible number. Using the bottom panel in Fig.~\ref{fig:streamvrgasdust1d} as a reference, the gas velocity has a positive correlation with local gas surface density. In the rings, the accretion stream has a radial velocity of up to $\approx150~{\rm m/s}$, more than half the local sound speed; it is comparable to that in the fast accretion stream shown in Fig.~3a of \cite{2018MNRAS.477.1239S}. Note that the inward radial speed of the dust can be higher than that of the gas, particularly in the outer halves of the dense gas rings (see Fig.~\ref{fig:streamvrgasdust1d}a). This is primarily because of the radially inward dust drift relative to the (infalling) gas in the sub-Keplerian regions, particularly the outer halves of gas rings. As shown in the velocity streamlines in Fig.~\ref{fig:rhostreamzoom}, the mass accretion flux is well concentrated in the stream area in most part of the disk. When the stream reaches a gap, the highly concentrated magnetic field lines force the accretion flux to follow the direction of the magnetic field, thus lifting the flux away from the midplane. This slows down the radial motion substantially, to as slow as $\approx50~{\rm m/s}$, or $\approx10\%$ of the local sound speed.

We next consider the radial dust-gas drift in the highly structured accretion stream. For reference, we note that in the 1D model of a smooth disk, this quantity is given by \citep[e.g.,][]{2010A&A...513A..79B}: 
\begin{equation}
    v_{\rm r,d}= - \frac{k_P(c_s/v_K)^2}{\tau+\tau^{-1}}v_K
    \label{eq:v_drift}
\end{equation}
where $k_P=d\ln{P}/d\ln{R}$ is the gas pressure gradient in log space, $R$ is the cylindrical radius, $v_K$ is the local Keplerian speed, and $\tau$ is the Stokes number of the dust. Supplying the disk parameters of our fiducial setup with initial conditions ($k_P=-2.5$), we get:
\begin{eqnarray}
   v_{\rm r,d}&=&-57.9\frac{a_{3.3}}{1+9.67\times10^{-6}a_{3.3}^2r_{\rm au}}{\rm cm~s^{-1}}\\ 
   &\simeq&-57.9\ a_{3.3}\ {\rm cm~s^{-1}} \nonumber
\end{eqnarray}
where $a_{3.3}$ is the dust size divided by 3.3 mm and $r_{\rm au}=r/{\rm au}$.
In Fig.~\ref{fig:kpvr_drift1d}, we find the gas pressure gradient $k_P$ in the accretion stream is quite different from a smooth disk: its absolute value at the representative time of $t=5000$~year of the fiducial simulation is almost one order of magnitude higher in some regions than in a smooth disk. The drift of the dust relative to the gas is sped up by the gas substructure because of its steeper pressure gradient compared to a smooth disk. 
With the measured $k_P$ (top panel of Fig.~\ref{fig:kpvr_drift1d}), we can calculate the dust radial drift velocity expected from Eq.~\ref{eq:v_drift}. The result is the orange solid line in the middle panel of Fig.~\ref{fig:kpvr_drift1d}. The good agreement between the direct measurement (``simulation'') and the calculation shows that the 1D model is still a powerful tool for analyzing the dust radial drift in a very dynamic disk. We should reiterate that the dust-gas drift speed ($\approx 10~{\rm m/s}$), although already much larger compared to a smooth disk because of the gas substructures, is still much smaller than the gas accretion speed in the midplane stream ($> 100~{\rm m/s}$ at many radii), as can be seen from a comparison of the middle panel of Fig.~\ref{fig:kpvr_drift1d} and the top panel of Fig.~\ref{fig:streamvrgasdust1d}.

\section{Parameter Exploration}
\label{sec:parameter}

\begin{figure*}
    \centering
    \includegraphics[width=1.0\textwidth]{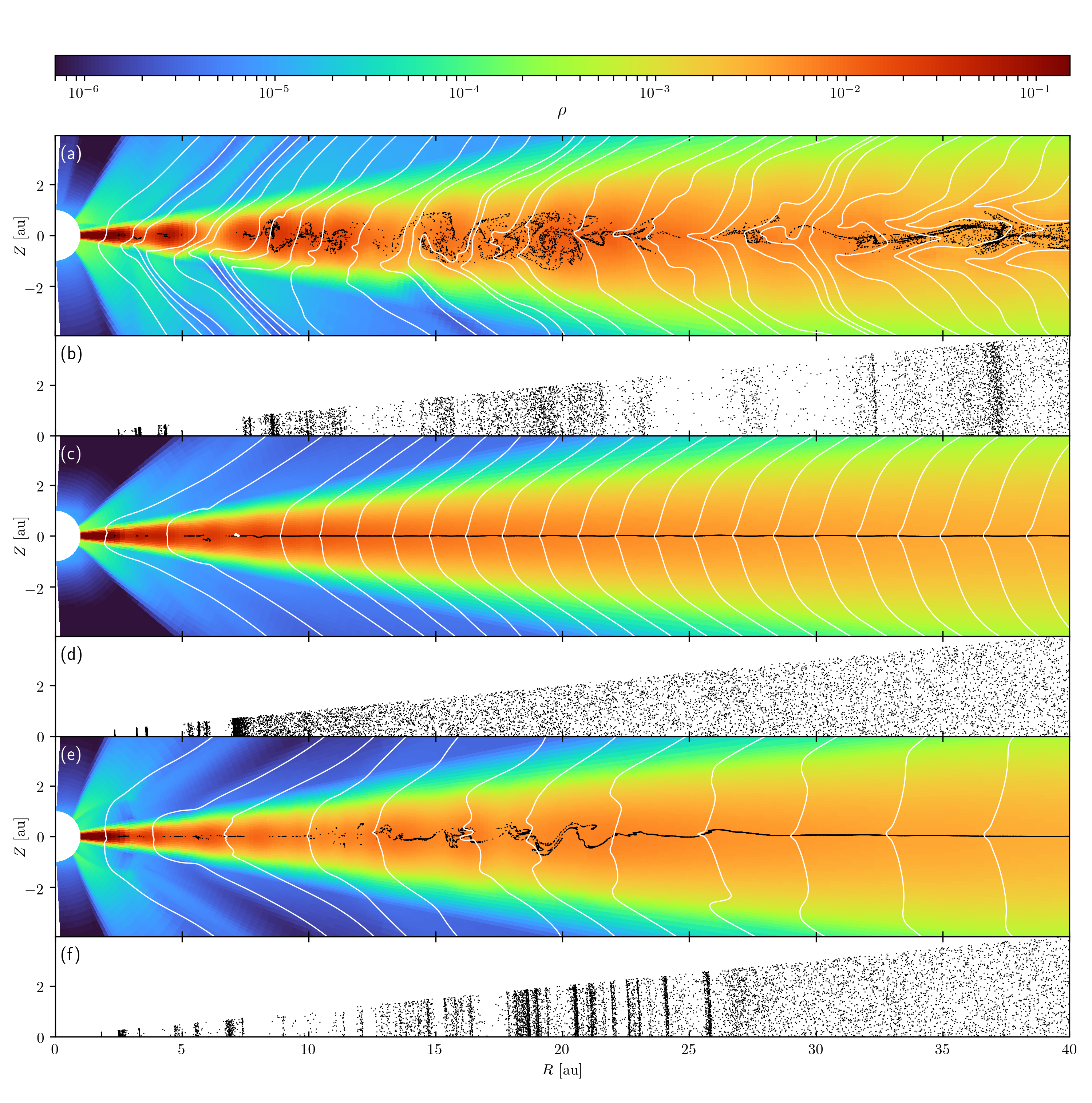}
    \caption{A summary of three models with different MHD parameters. Note we plotted with smaller radial extent (0 to 40 ~au) than similar figures of the fiducial model. Panels (a) and (b) show, respectively, the edge-on and face-on view of the better magnetically coupled model (with $Am_0=0.75$) at t=1200 years. Panels (c) and (d) show the more magnetically diffusive model (with $Am_0=0.08$) at t=5000 years. Panels (e) and (f) show the weaker field model (with initial $\beta=5000$) at t=5000 years. For panels (a), (c) and (e), the background color map is for the gas density, and the white lines show the poloidal field lines. For simplicity, only 3.3 mm particles are plotted (the black dots). (See the supplementary material in the online journal for an animated version of this figure for the weak-field case.) 
    }
    \label{fig:para_rho}
\end{figure*}

In this section we briefly explore the effects of varying the magnetic diffusion coefficient and magnetic field strength. The results are summarized in Fig.~\ref{fig:para_rho}. 

For the magnetic diffusion, we only change the scaling of ambipolar Elsasser number, $Am_0$ in Eq.~\ref{eq:am}. With $Am_0=0.75$, three times of the fiducial case, the gas and magnetic field have a better coupling. The substructures evolve more quickly, which is to be expected. Prominent rings and gaps have already formed up to 40~au at t=1200 years in both the gas (Fig.~\ref{fig:para_rho}a) and dust (Fig.~\ref{fig:para_rho}b). Dust particles are strongly stirred up vertically, similar to our fiducial case but at much later times (t=5000 years). The more magnetically diffusive case is shown in the panels (c) and (d)  of Fig.~\ref{fig:para_rho}. Here $Am_0=0.08$, three times smaller than the fiducial case. The gas disk is much more quiescent, with little obvious substructure visible even after t=5000 years of evolution. This is not surprising because the higher magnetic diffusivity makes it more difficult for the field lines to bend and exert a force on the gas. Lacking vertical gas motion, dust particles settle quickly to the midplane, forming a thin and flat sheet over most of the disk. This set of simulations with different magnetic diffusivities demonstrates that the formation of gas and dust substructures in a relatively strongly  magnetized ($\beta=10^3$) disk is robust unless the field becomes too weakly coupled to the gas.

\begin{figure}
    \centering
    \includegraphics[width=0.48\textwidth]{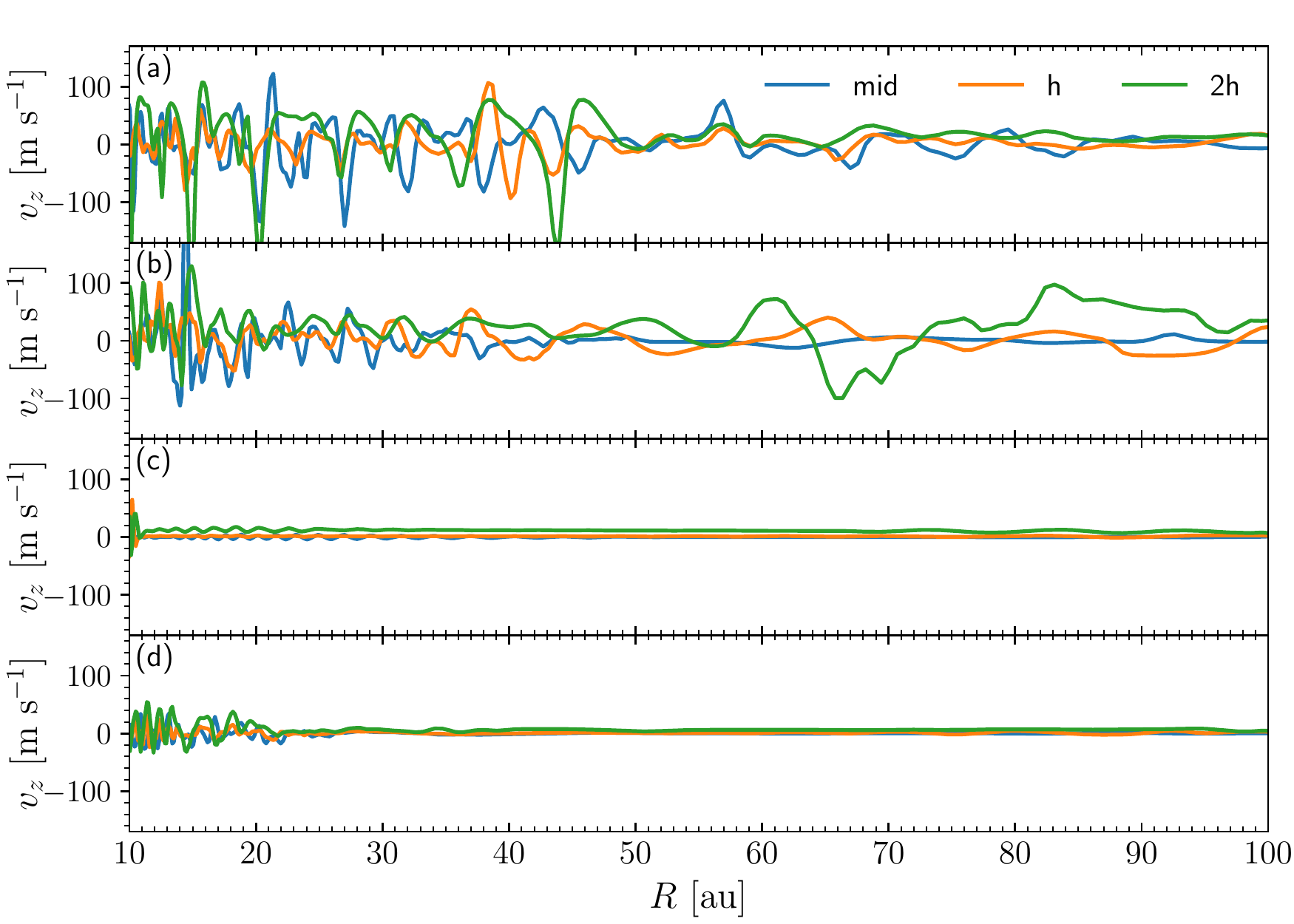}
    \caption{A comparison of gas vertical motion between four models. The three lines of different colors are $v_z$ measured at different height above midplane: ``mid'' means measurement at midplane, ``h'' means one scale height above midplane, and ``2h'' means two scale heights. From top to bottom, panel (a), (b), (c), and (d) shows, respectively, our fiducial model, model with $Am_0=0.75$, $Am_0=0.08$ and $\beta=5000$. All measurements are taken at t=5000 years except for panel (b), which is 1200 years. The same legend is shared by all panels.}
    \label{fig:para_vz}
\end{figure}

Compared to those of the more and less magnetically diffusive cases, the results of the weaker field case of $\beta=5000$ are more unexpected, especially in terms of the dust substructure formed. For example, at the time shown in Fig.~\ref{fig:para_rho}e,f (t=5000 years), the particles are still stirred up by the meridional flows within a radius of $\sim 30$~au, but the vertical extent of the dust distribution (Fig.~\ref{fig:para_rho}e) is smaller than that in the fiducal model (Fig.~\ref{fig:dust_meridian}c) or its better coupled counterpart (Fig.~\ref{fig:para_rho}a). The smaller vertical extent comes from the fact that, although vigorous magnetically-driven meridional flows are still present (see a movie version of Fig.\ref{fig:para_rho} with the weaker field case only in the supplementary material), they are significantly slower than those in the fiducial case (see Fig.~\ref{fig:para_vz} below) because of a weaker initial magnetic field. As a result, the dust particles can settle more easily in the vertical direction. 
One might expect the slower meridional flows to produce less substructure in the dust surface distribution, but this does not appear to be the case: we see numerous well-defined rings and gaps that are formed inside about 25~au at the time shown in Fig.~\ref{fig:para_rho}f. The reason is that most of the dust is concentrated vertically near the fast midplane accretion stream where the flow is especially dynamic and spatially variable, which makes the dust distribution highly inhomogeneous and time-dependent as well. The structuring of the dust distribution by a dynamic gas flow is consistent with the fact that the dust radial velocity is highly correlated with the gas radial velocity, 
as in the fiducial case (see Fig.~\ref{fig:streamvrgasdust1d}). 
Compared to the (more strongly magnetized) fiducial case, the meridional flows are aided to a larger extent by surface accretion streams (termed ``avalanche accretion stream'' by \citet{2017MNRAS.468.3850S}; see their Fig.~4 for an example; see also \citet{ZhuStone2018}), which slide down the disk surface towards the center due to angular momentum loss from magnetic braking. Part of the lost angular momentum is transported along magnetic field lines to the disk midplane region, where it drives outward expansion (see this phenomenon in the movie version of Fig.~\ref{fig:para_rho}e,f in the supplementary materials). 
The weaker field simulation strengthens the case for dust substructure formation in magnetized disks as long as the field remains sufficiently well coupled to the gas.

To quantify the meridional flows, we compare the strengths of the vertical motions of all four models discussed in this paper (including the fiducial model) in Fig.~\ref{fig:para_vz}. We find that, even at the midplane, the gas vertical velocity is around 40~m/s in most radii for the fiducial model, and exceeds 100~m/s at several locations. From the other two (orange and green) lines in panel (a), we find the gas vertical motion generally gets stronger as it goes up and above the midplane. The only case that has a similar level of vertical stirring is the better magnetically coupled case (panel b, with $Am_0=0.75$). The more magnetically diffusive model has very little vertical motion, except near the inner radial boundary, which is the reason for the small thickness of its settled dust layer (see Fig.~\ref{fig:para_rho}c). For the weaker initial magnetic field case ($\beta=5000$), the vertical motions are not as strong as the fiducial case, but are still significant in the inner part of the disk (within about 25 au of the central star) where most of the dust rings and gaps reside. This is consistent with the picture that the meridional gas flows are important in shaping the dust substructure.

\section{Discussion}
\label{sec:discussion}

In our simulations, dust substructure naturally develops in the highly structured wind-launching non-ideal MHD gas disks, as found previously \citep[e.g.,][]{2020A&A...639A..95R}. However, the dust substructure formation turns out to be much more subtle and richer than anticipated. Specifically, the radial dust distribution is driven primarily by the gas advection rather than the (pressure gradient-driven) non-Keplerian gas motions 
envisioned in the traditional mechanism. Associated with our new picture of dust distribution are several unique features of both gas and dust that may be tested by observations.

Firstly, there is a fast (corrugated) midplane gas accretion stream that is essential for dust radial transport. The stream is a robust gas feature physically caused by the strong magnetic braking associated with the toroidal magnetic field reversal. It serves as both a trap and a conveyor belt for the grains, particularly the large ones, that enter it through either gravitational settling or meridional converging flow. The trapped grains move rapidly inward, through both gas advection and radial drift due to sub-Keplerian gas rotation caused by strong magnetic braking. The average accretion speed in the narrow stream is of order $100{\rm ~m~s^{-1}}$ for our reference simulation, which has a mass accretion rate of order $10^{-6}~M_\odot~{\rm yr}^{-1}$ that is more appropriate for the youngest, deeply embedded protostars \citep[e.g.,][]{2017ApJ...834..178Y}. Such a large speed is in principle detectable with ALMA. However, the detectability will depend on the availability of suitable molecules that trace the disk midplane and the extent to which the molecule emission is attenuated by opacity in both line and continuum. 

Secondly, the dust grains are {\it not} concentrated at the density maxima inside gas rings, as one would expected if the dust concentration was dominated by pressure gradient-induced radial drift. Rather, they are mostly concentrated in the inner half of the rings (see Figs.~\ref{fig:dust_faceon} and \ref{fig:dust_meridian}) where the gas rotation is predominantly super-Keplerian (see Fig.~\ref{fig:sub-super-K}). The reason for the offset is that the dust concentration in our simulations is dominated by the gas kinematics rather than its pressure gradient based on the fact that the dust velocity is typically dominated by the gas velocity rather than the dust-gas drift velocity (see the lower left panels of Fig.~\ref{fig:vrvz_drift} and associated discussion). The implication is that the dust continuum emission should peak interior to the gas density maxima. This radial offset between the dust and gas distributions is a telltale sign of our gas advection-dominated dust concentration. It should be searched for observationally, especially in the most actively accreting, earliest Class 0 disks. The dust distribution should be relatively easy to measure through high-resolution dust continuum emission, through projects such as the ALMA Large Program ``eDisk'' (2019.1.0026.L, PI: N. Ohashi). Measuring the gas distribution at a comparable resolution would be more challenging, and may be limited by optical depth effects and chemical variation. Alternatively, we can use the deviation the gas rotation from the local Keplerian speed as a proxy for the gas distribution \citep[e.g.,][]{2019Natur.574..378T}. We note that some of the dust rings in the face-on view are significantly narrower than the gas rings (see Fig.~\ref{fig:dust_faceon}), although an observational test of the difference will again be limited by the difficulty in measuring the gas distribution.

Perhaps most excitingly, our fiducial model has strong meridional circulation motions that may be readily observable. 

\subsection{A New Scenario for Disk Meridional Circulation}

One of the most exciting recent advances in protoplanetary disk research is the detection of meridional motions from ALMA molecular line observations. In the best studied case, HD 163296, \citet{2021ApJS..257...18T} inferred the presence of meridional flows with speeds of tens of m/s (see their Fig.~5 and also \citealt{2019Natur.574..378T}). One attractive interpretation is that these flows are generated by embedded planets \citep{2016ApJ...832..105F,2021MNRAS.502.5325R}, which, if massive enough, open up gaps in the gas disk and stir up meridional motions near the inner and outer edges of the gaps \citep{2021MNRAS.502.5325R}. Another interpretation is that these flows are potentially produced by vertical shearing instability (VSI), as demonstrated by, e.g., \citet{2021A&A...653A.113B}. Here, we propose an entirely different scenario where the meridional motions are driven by magnetic fields in a weakly ionized disk through non-ideal MHD effects.  

\begin{figure*}
    \centering
    \includegraphics[width=1.0\textwidth]{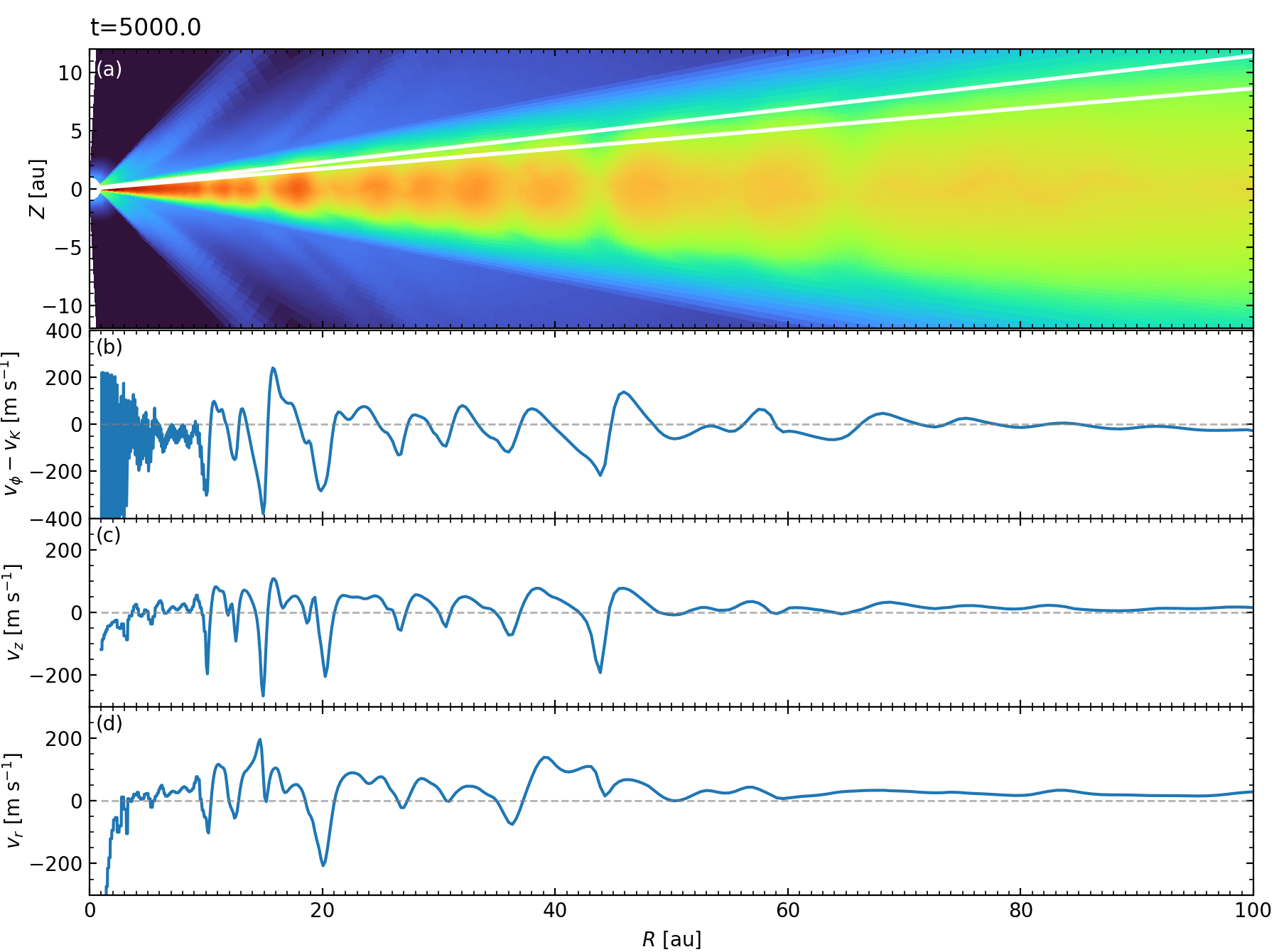}
    
    \caption{Gas kinematics near the disk surface. Panel (a) shows the density color map in a meridional plane, with the white lines marking the region where the gas velocities are averaged. Also plotted are the derviation of gas rotation speed from the local Keplerian speed (panel b), the vertical velocity component (c), and the radial velocity component (d).  }
    \label{fig:gas3v}
\end{figure*}

The rigorous meridional flows in our fiducial non-ideal MHD simulation can be seen most vividly in Fig.~\ref{fig:LIC} (a movie version is available in the supplementary materials), where the gas velocity field in the meridional plane is displayed through line integral convolution (LIC). It can also be seen in Fig.~\ref{fig:rhostreamzoom}, where several representative streamlines with velocity directions are plotted in a region that contains prominent rings and gaps. To be more quantitative, we plot in Fig.~\ref{fig:gas3v} the meridional velocities $v_r$ and $v_z$ and the deviation of the rotation speed from the local Keplerian speed near the disk surface (two scale heights above the midplane), averaged over a strip of half scale height in width. Clearly, the meridional flow speeds readily reach the observationally inferred range of tens of m/s; if anything, they may be somewhat higher than observed (up to 200 m/s). This plot demonstrates that the new magnetic field-based mechanism can generate meridional flows that are fast enough to be compatible with the current observations. 

\begin{figure*}
    \centering
    \includegraphics[width=1.0\textwidth]{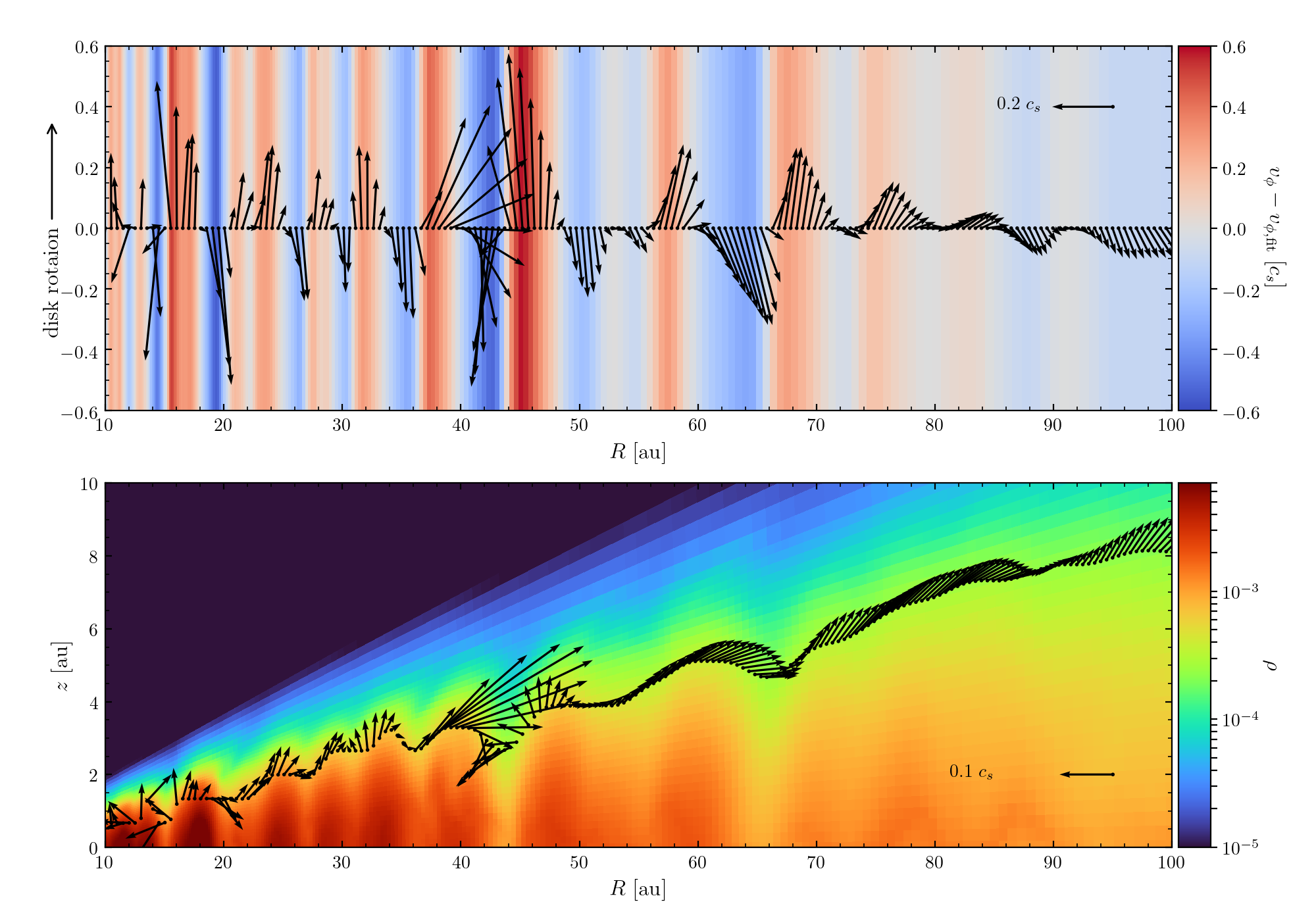}
    \caption{Gas velocity structure presented the same general style as in Fig.~2 of \citet{2019Natur.574..378T} to facilitate comparison with observations. {\bf Upper Panel:} Red and blue regions have, respectively, faster and slower rotation speeds relative to the best fit power-law profile of the gas azimuthal velocity at a representative surface, $v_{\phi, {\rm fit}}$, with the black arrows showing the velocity vectors in the radial and azimuthal directions. {\bf Lower Panel:} Gas velocity vectors in the meridional plane at the surface, with a vector for 0.1~$c_s$ shown for reference. Plotted in the background color map is the gas density, highlighting the rings and gaps.}
    \label{fig:obs_vector}
\end{figure*}


To make a closer contact with observations, we follow \citet{2019Natur.574..378T} and plot in Fig.~\ref{fig:obs_vector} the gas velocity structure in $R-\phi$ and $R-z$ plane, where $R$ is the cylindrical radius and $z$ the height above the midplane. For illustration purposes, we picked a surface that has a constant column density above it. 
This figure is to be compared with the Fig.~2 of \citet{2019Natur.574..378T} for the HD 163296 disk. Since our simulation was not specifically tailored for the HD 163296 system, a detailed match is not to be expected. Nevertheless, 
there are some broad similarities. For example, there are at least three so-called ``collapsing'' regions (where the meridional flow points toward the midplane) in the $R-z$ plane, at $\sim 15$, 21, and 44 au, respectively. The (downward) ``collapsing'' speed is of order 10\% of the local sound speed, comparable to observationally inferred values. The collapsing region at 44 au is particularly similar to the $\mathrm{CO}$ flow structure in HD 163296, which has a gap surrounded by super and sub-keplerian rotation. The circulation pattern in the gap can explain this similarity. In Fig.~\ref{fig:rhostreamzoom}, we show the streamlines originated near the center of the ring at 45-55~au. After the accretion stream is diverted away from the midplane by the strong poloidal magnetic field near the inner edge of the ring, it can move in one of three directions: (a) going back towards the midplane and rejoin the accretion stream that enters into the next (interior) ring; (b) following the open poloidal field lines into the disk wind; and (c) turning back outward and then going down to the midplane, rejoining the accretion stream in the same ring, but near the outer edge. Component (a) mainly exists in the gap, where the relatively low density of the gas makes the velocity field pointing towards the midplane relatively easy to trace using relatively abundant molecules such as CO. Component (b) is near the wind base, so the observed velocity field will be sensitive to the physical condition of the wind launching. 
Component (c) may be harder to detect if it is buried deep inside the dense ring. However, when the gas flow reaches the ring's outer edge, where the vector points back down, the lower gas density there makes it more easily observable again. 

In addition to the collapsing regions, our fiducial simulation also produces alternating radial zones of sub- and super-Keplerian rotation (see the upper panel of Fig.~\ref{fig:obs_vector}), which are reminiscent of the pattern inferred for HD 163296 (see the upper panel of Fig.~2 of \citealt{2019Natur.574..378T}). Perhaps more importantly, our simulation shows a clear disk-wind feature in the outer disk, which is also inferred in HD 163296. This is important because the (magnetic) disk-wind is an integral part of the weakly magnetized, weakly ionized disk system, and plays an important part in creating the rings and gaps as well as the meridional flows. In other words, unlike the interpretation of the collapsing flows involving embedded planets that requires a separate mechanism to explain the signature of a disk-wind, our mechanism can naturally account for both the collapsing flows and the disk wind. Indeed, there are independent lines of evidence for a disk-wind in HD 163296 from high-resolution ALMA CO (2-1) molecular line observations at millimeter \citep{2021ApJS..257...16B} and  infrared CO ro-vibrational lines \citep[e.g.,][]{2016MNRAS.458.1466H}, which strengthen our interpretation.

There is another generic difference between the meridional flow patterns induced by an embedded planet and our proposed mechanism that is crucial for distinguishing between the two. In the former case, the midplane gas flows outwards away from the planet-induced gap, driving a meridional gas circulation in the upper half of the outer edge of the gap that is counter-clockwise (see Fig.~6 of \citealt{2016ApJ...832..105F}). This is the opposite of our case, where the gas flow near the midplane is always directed inwards, which drives a clock-wise meridional gas circulation in the dense ring outside the gap (see Fig.~\ref{fig:rhostreamzoom}). 

\begin{figure}
    \centering
    \includegraphics[width=0.48\textwidth]{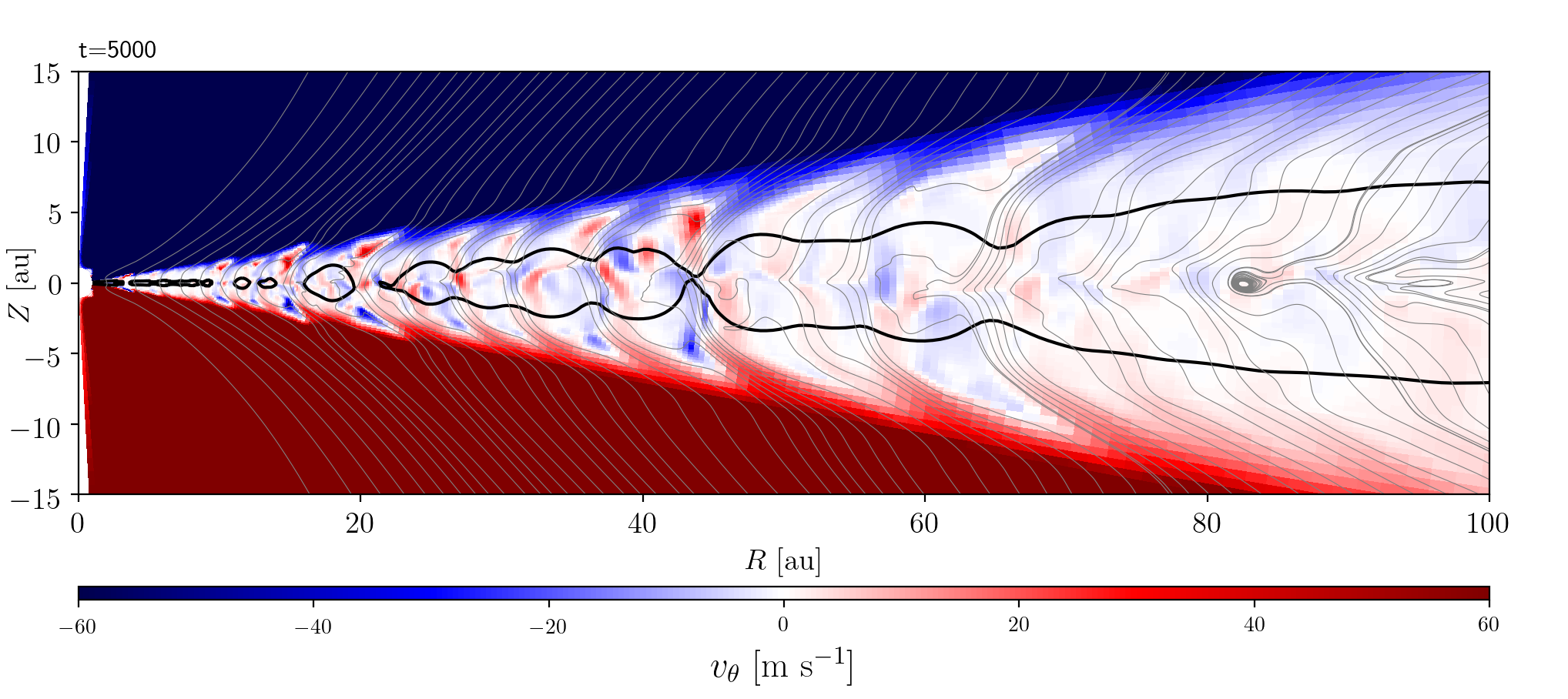}
    \caption{Spatial distribution of the polar component of the meridional flow, $v_\theta$, for the fiducial model at the representative time $t=5000$~years. Note that the meridional flow converges towards or diverges from the midplane in a broadly mirror symmetric pattern, which is the opposite of the pattern produced by vertical shear instability (contrast this figure with Fig.~3 of \citealt{2021A&A...653A.113B}). The unit for the numbers on the colorbar is m~s$^{-1}$. The thin grey lines are polodial field lines and the thick black line is an iso-density (normalized by the initial density) contour highlighting the gas rings and gaps.}
    \label{fig:vtheta}
\end{figure}

The meridional flow pattern found in our simulation is also different from that induced by VSI. For example, in the VSI simulation displayed in Fig.~3 of \citet{2021A&A...653A.113B}, the vertical component of the meridional flow is continuous when the gas crosses the disk midplane from one hemisphere to the other (see their third panel in the bottom row). This is the opposite of that seen in our simulation where the meridional gas circulation is driven primarily by the fast accretion stream near the midplane, which is diverted upwards in the upper hemisphere and downwards in the lower one, forming a pattern that is broadly mirror symmetric, as illustrated in Fig.~\ref{fig:vtheta}, which plots the spatial distribution of the $\theta-$ component of the meridional flow $v_\theta$. Another difference is that we find alternating vertical stripes of sub- and super-Keplerian regions that are continuous across the disk midplane (see Fig.~\ref{fig:sub-super-K}), whereas the sub-Keplerian region flips to become super-Keplerian across the midplane and vise versa for the VSI-induced meridional flow pattern (see the bottom-right panel of Fig.~3 of \citealt{2021A&A...653A.113B}). Whether these different velocity patterns with respect to the midplane can be observed or not depends how well the velocity field in the hemisphere of the disk facing away from the observer can be probed by molecular lines. 

\subsection{Future refinements}

There are several future directions for the line of research pursued in this investigation. One of the most obvious improvements would be to include the dust feedback on the gas dynamics, which is expected to be particularly significant in regions where the dust is concentrated \citep[e.g.,][]{2018ApJ...868...27Y,2020A&A...635A.190S,2022ApJ...924....3X}. Another improvement is to include a more detailed calculation of the non-ideal MHD coefficients and to include other non-ideal effects (especially the Hall effect) in addition to ambipolar diffusion. Furthermore, it would be interesting to see whether the vigorous meridional gas motions persist in fully three dimensional simulations and whether such motions concentrate dust particles in a way broadly similar to what we found in our 2D (axisymmetric) simulations. It would also be interesting to explore the implications that our new mechanism of dust concentration through advection by meridional gas flows may have on grain growth. 

\section{Conclusion}
\label{sec:conclusion}

We have carried out 2D (axisymmetric) MHD disk simulations including ambipolar diffusion and Lagrangian dust particles of three different sizes (1~mm, 3.3~mm, and 1~cm). We focused on the fiducial case, with an initial field strength corresponding to a plasma-$\beta=10^3$ on the disk midplane and an AD coefficient characterized by an Elsasser number $Am_0=0.25$. We also briefly explored the gas and dust dynamics in three other models with different values of $\beta$ and $Am_0$. Our main conclusions are as follows.

(1) Prominent rings and gaps spontaneously develop in the gas distribution, strengthening the case for substructure formation in non-ideal MHD disks found in previous simulations.
We found a characteristic meridional flow pattern 
driven by a fast accretion stream near the midplane 
(see Fig.~\ref{fig:LIC} and \ref{fig:rhostreamzoom}). The accretion stream is diverted away from the midplane at the inner edge of a dense ring by a strong poloidal magnetic field, part of which crosses the field lines and converges  back towards the midplane to form the field-crossing funnel flow. Part of the remaining diverted flow leaves the disk surface in a wind, with the rest moving outward to form a back-circulation that carries the gas to the outer part of the ring where it rejoins the midplane accretion stream. 

(2) The meridional gas flow pattern is the key to the dust transport in our simulation, where prominent rings and gaps are present in the face-on view of the dust distribution
(see Fig.~\ref{fig:dust_faceon}). Although the grains have a general tendency to drift radially relative to the gas towards the centers of gas rings, 

their spatial distribution is primarily controlled by the gas motions, which are typically much faster than the dust-gas drift in the meridional plane (see Fig.~\ref{fig:vrvz_drift}). In particular, the grains that have settled near the midplane are carried rapidly inwards by the fast accretion stream to the inner edges of dense rings, where they are lifted up by the gas flows diverted away from the midplane. A large portion of the lifted-up grains are advected radially outward by the gas back circulation, but they settle back towards the midplane before reaching the outer parts of the rings, especially for the larger grains. This creates a puffed-up dust concentration in the inner parts of the gas rings (where the gas rotation is predominantly super-Keplerian, see Fig.~\ref{fig:sub-super-K}) that may be tested through high-resolution continuum and line observations. 

(3) The flow pattern in our simulation provides an attractive explanation for the meridional flows recently inferred in HD 163296 and other disks \citep[e.g.,][]{2021ApJS..257...18T}. In particular, we found several ``collapsing'' regions in our fiducial simulation where the gas near the disk surface converges towards the midplane (see Fig.~\ref{fig:obs_vector}), with speeds comparable to those inferred in the HD 163296 disk. 
The meridional velocity pattern is very different from that induced by embedded planets or the vertical shear instability, and may be observationally distinguishable from these other mechanisms. One advantage of our mechanism is that it naturally explains the disk-wind signature inferred in the outer part of the HD 163296 disk, which would require a separate explanation in the other mechanisms. 

(4) The substructure formation in both the gas and dust in a non-ideal MHD disk depends on the degree of magnetic coupling and the strength of the magnetic field (see Fig.~\ref{fig:para_rho}). A better magnetic coupling enables an earlier development of the rings and gaps in the gas, as well as stronger meridional motions that create more dust substructure. 
The meridional gas motions become slower for a weaker magnetic field, although significant dust substructure can still develop if the field is sufficiently well coupled to the gas.

\section*{Acknowledgements}

We thank the referee for detailed and constructive comments, which improved the presentation of the paper. XH acknowledges support from the University of Virginia through VICO (Virginia Initiative on Cosmic Origins) and NSF AST-1716259.
ZYL is supported in part by NASA 80NSSC20K0533 and NSF AST-1910106.
CCY, ZZ, and XH are grateful for the support from NASA via the Astrophysics Theory Program (grant number 80NSSC21K0141).
CCY and ZZ also thanks the support from NASA via the Emerging Worlds program (grant number 80NSSC20K0347).
ZZ acknowledges support from the National Science Foundation under CAREER Grant Number AST-1753168.
CCY also acknowledges the support from NASA via the Theoretical and Computational Astrophysics Networks program (grant number 80NSSC21K0497).
Our simulations are made possible by an XSEDE allocation (AST200032). 
\section*{Data Availability}

The data from the simulations will be shared on reasonable request to the corresponding author.



\bibliographystyle{mnras}
\bibliography{ref,ref_mhd_dust} 







\bsp	
\label{lastpage}
\end{document}